\documentclass[11pt,a4paper]{article}

\usepackage{jheppub}

\RequirePackage{ifpdf}
\usepackage[dvipsnames]{xcolor}
\usepackage{amsmath}
\usepackage{amsfonts,amscd,dsfont}
\usepackage{feynmp}
\usepackage{amsthm}
\usepackage{amssymb}
\usepackage{epsfig,psfrag}
\usepackage[toc,page]{appendix}
\usepackage{etoolbox}
\usepackage{graphicx}

\usepackage{float}

\RequirePackage{tikz}
\usetikzlibrary{arrows,calc,shapes.misc,decorations.markings,snakes}
\tikzset{cross/.style={cross out, draw=black, minimum size=2*(#1-\pgflinewidth), inner sep=0pt,
        outer sep=0pt},cross/.default={3pt},
    gluon/.style={decorate, decoration={coil,aspect=0.9,segment length=5pt, amplitude=3pt}}}

\usepackage{axodraw}
\usepackage{subfig}

\def\be{\begin{equation}}
\def\ee{\end{equation}}
\newcommand{\beq}{\begin{equation}}
\newcommand{\eeq}{\end{equation}}
\newcommand{\bea}{\begin{align}}
\newcommand{\eea}{\end{align}}
\newcommand{\bsp}{\begin{equation}\begin{split}}
\newcommand{\esp}{\end{split}\end{equation}}
       \def\beq{\begin{equation}}
\def\eeq{\end{equation}}
\def\bsp#1\esp{\begin{split}#1\end{split}}

\newcommand{\cE}{\begin{cal}E\end{cal}}
\newcommand{\cF}{\begin{cal}F\end{cal}}
\newcommand{\cG}{\begin{cal}G\end{cal}}
\newcommand{\cH}{\begin{cal}H\end{cal}}
\newcommand{\cI}{\begin{cal}I\end{cal}}

\newcommand{\cN}{\begin{cal}N\end{cal}}
\newcommand{\cO}{\begin{cal}O\end{cal}}

\newcommand{\cR}{\begin{cal}R\end{cal}}

\newcommand{\fa}{\mathfrak{a}}
\newcommand{\fb}{\mathfrak{b}}
\newcommand{\fc}{\mathfrak{c}}
\newcommand{\fd}{\mathfrak{d}}

\newcommand{\zbar}{{\bar z}}
\newcommand{\zb}{{\bar z}}
\newcommand{\bx}{\textbf{x}}
\newcommand{\bp}{\textbf{p}}

\newcommand{\Res}{\textrm{Res}}

\title{The Multi-Regge Limit of the Eight-Particle Amplitude Beyond Leading Logarithmic Accuracy
}

\author[a]{Robin Marzucca}
\author[a]{Bram Verbeek}

\affiliation[a]{Center for Cosmology, Particle Physics and Phenomenology (CP3), \\
UCLouvain, Chemin du Cyclotron 2, 1348 Louvain-La-Neuve, Belgium.
}

\preprint{CP3-18-63}

\abstract{
We present the computation of the eight-particle three-loop amplitude beyond leading logarithmic accuracy in the multi-Regge limit of planar $\cN=4$ Super Yang-Mills theory. Starting from the all-loop dispersion integral form of the amplitude, we consider the eight-particle case and by analyzing said dispersion integral we associate it to a well-defined Fourier-Mellin transform. By using the properties of the Fourier-Mellin representation and its convolution product structure, we compute the three-loop eight-particle MHV amplitude at next-to-leading logarithmic accuracy. From this MHV result, we obtain the three-loop eight particle amplitude in multi-Regge kinematics for all helicity configurations, including next-to-next-to-MHV.
Finally, we find that the result is described by combinations of single-valued multiple polylogarithms of uniform weight, the leading singularity structure of which corresponds to the classification shown at leading logarithmic accuracy.
}


\newcommand{\contourThree}{
\begin{tikzpicture}{scale=0.75, every node/.style={transform shape}}
        \draw[->,thick] (0,3) -- (6,3) node [right] {$\Re(\nu_1)$};
        \draw[->,thick] (3,0) -- (3,6) node [above] {$\Im(\nu_1)$};
        \draw (4.5,3) node[cross] (A) {};
        \node [below of=A, node distance=10pt] {$\pi \Gamma$};
        \draw (1.5,3) node[cross] (mA) {};
        \node [below of=mA, node distance=10pt] {$\nu_{2}$};

        \draw[thick,RoyalBlue,yshift=2, decoration={
                markings,
                mark = at position 0.23 with {\arrow{>}},
                mark = at position 0.4 with {\arrow{>}},
                mark = at position 0.6 with {\arrow{>}},
                mark = at position 0.8 with {\arrow{>}}},
                postaction={decorate}]
        (0,3) -- (1.2, 3) arc (-180:0:.3) -- (4.2, 3) arc (180:0:.3) -- (6,3);

\end{tikzpicture}}

\newcommand{\contourTwo}{
\begin{tikzpicture}{scale=0.75}
        \draw[->,thick] (0,3) -- (6,3) node [right] {$\Re(\nu_{2})$};
        \draw[->,thick] (3,0) -- (3,6) node [above] {$\Im(\nu_{2})$};
        \draw (4.5,3) node[cross] (A) {};
        \node [below of=A, node distance=10pt] {$\nu_{1}$};
        \draw (1.5,3) node[cross] (mA) {};
        \node [below of=mA, node distance=10pt] {$\nu_{3}$};

        \draw[thick,RoyalBlue,yshift=2, decoration={
                markings,
                mark = at position 0.23 with {\arrow{>}},
                mark = at position 0.4 with {\arrow{>}},
                mark = at position 0.6 with {\arrow{>}},
                mark = at position 0.8 with {\arrow{>}}},
                postaction={decorate}]
        (0,3) -- (1.2, 3) arc (-180:0:.3) -- (4.2, 3) arc (180:0:.3) -- (6,3);

\end{tikzpicture}}

\newcommand{\contourOne}{
\begin{tikzpicture}{scale=0.75}
        \draw[->,thick] (0,3) -- (6,3) node [right] {$\Re(\nu_{3})$};
        \draw[->,thick] (3,0) -- (3,6) node [above] {$\Im(\nu_{3})$};
        \draw (4.5,3) node[cross] (A) {};
        \node [below of=A, node distance=10pt] {$\nu_{2}$};
        \draw (1.5,3) node[cross] (mA) {};
        \node [below of=mA, node distance=10pt] {$-\pi \Gamma$};

        \draw[thick,RoyalBlue,yshift=2, decoration={
                markings,
                mark = at position 0.23 with {\arrow{>}},
                mark = at position 0.4 with {\arrow{>}},
                mark = at position 0.6 with {\arrow{>}},
                mark = at position 0.8 with {\arrow{>}}},
                postaction={decorate}]
        (0,3) -- (1.2, 3) arc (-180:0:.3) -- (4.2, 3) arc (180:0:.3) -- (6,3);

\end{tikzpicture}}

\newcommand{\contourThreeShifted}{
\begin{tikzpicture}{scale=0.75, every node/.style={transform shape}}
        \draw[->,thick] (0,3) -- (6,3) node [right] {$\Re(\nu_1)$};
        \draw[->,thick] (3,0) -- (3,6) node [above] {$\Im(\nu_1)$};
        \draw (4.5,3) node[cross] (A) {};
        \node [below of=A, node distance=10pt] {$\pi \Gamma$};
        \draw (1.5,3) node[cross] (mA) {};
        \node [below of=mA, node distance=10pt] {$\nu_{2}$};

        \draw[thick,RoyalBlue,yshift=2, decoration={
                markings,
                mark = at position 0.23 with {\arrow{>}},
                mark = at position 0.4 with {\arrow{>}},
                mark = at position 0.6 with {\arrow{>}},
                mark = at position 0.8 with {\arrow{>}}},
                postaction={decorate}]
        (0,3) -- (1.2, 3) arc (-180:0:.3) -- (4.2, 3) arc (-180:0:.3) -- (6,3);

\end{tikzpicture}}

\newcommand{\contourTwoShifted}{
\begin{tikzpicture}{scale=0.75}
        \draw[->,thick] (0,3) -- (6,3) node [right] {$\Re(\nu_{2})$};
        \draw[->,thick] (3,0) -- (3,6) node [above] {$\Im(\nu_{2})$};
        \draw (4.5,3) node[cross] (A) {};
        \node [below of=A, node distance=10pt] {$\nu_{1}$};
        \draw (1.5,3) node[cross] (mA) {};
        \node [below of=mA, node distance=10pt] {$\nu_{3}$};

        \draw[thick,RoyalBlue,yshift=2, decoration={
                markings,
                mark = at position 0.23 with {\arrow{>}},
                mark = at position 0.4 with {\arrow{>}},
                mark = at position 0.6 with {\arrow{>}},
                mark = at position 0.8 with {\arrow{>}}},
                postaction={decorate}]
        (0,3) -- (1.2, 3) arc (180:0:.3) -- (4.2, 3) arc (180:0:.3) -- (6,3);

\end{tikzpicture}}

\newcommand{\contourOneShifted}{
\begin{tikzpicture}{scale=0.75}
        \draw[->,thick] (0,3) -- (6,3) node [right] {$\Re(\nu_{3})$};
        \draw[->,thick] (3,0) -- (3,6) node [above] {$\Im(\nu_{3})$};
        \draw (4.5,3) node[cross] (A) {};
        \node [below of=A, node distance=10pt] {$\nu_{2}$};
        \draw (1.5,3) node[cross] (mA) {};
        \node [below of=mA, node distance=10pt] {$-\pi \Gamma$};

        \draw[thick,RoyalBlue,yshift=2, decoration={
                markings,
                mark = at position 0.23 with {\arrow{>}},
                mark = at position 0.4 with {\arrow{>}},
                mark = at position 0.6 with {\arrow{>}},
                mark = at position 0.8 with {\arrow{>}}},
                postaction={decorate}]
        (0,3) -- (1.2, 3) arc (-180:0:.3) -- (4.2, 3) arc (-180:0:.3) -- (6,3);

\end{tikzpicture}}

\begin{document}
\maketitle

\section{Introduction}

The planar limit of $\cN=4$ super Yang-Mills (SYM) theory is a rich setting for the study of scattering amplitudes, in which a myriad of interesting results has been uncovered in recent years. Beyond the ordinary conformal symmetry exhibited by $\cN=4$ SYM, the theory has a dual conformal symmetry \cite{Drummond:2006rz,Bern:2006ew,Bern:2007ct,Alday:2007hr,Drummond:2007aua,Brandhuber:2007yx} in the planar limit which closes with the aforementioned conformal symmetry to an infinite-dimensional Yangian symmetry \cite{Drummond:2009fd}. This infinite-dimensional symmetry is connected to integrability \cite{Beisert:2017ie}, suggesting it is possible to study the theory beyond the perturbative regime. The high degree of symmetry also has consequences for the structure of scattering amplitudes, for example, the four and five particle cases are completely determined by the so-called Bern-Dixon-Smirnov (BDS) ansatz \cite{Bern:2005iz}. This is no longer the case beyond five particles, where one finds an additional non-trivial dual conformally invariant contribution refered to as a remainder function \cite{Bern:2008ap}. Furthermore, the kinematic dependence of the amplitude can be written in terms of momentum twistors \cite{Hodges:2009hk} which reduces the kinematics to a configuration of points in three dimensional projective space \cite{Golden:2013xva}. This in turn suggests that scattering amplitudes are iterated integrals of one-forms on this space of configurations, the singularities of which are described by a mathematical structure called a cluster algebra. Furthermore, for maximal helicity violating (MHV) and next-to-maximal helicity violating (NMHV) amplitudes, these iterated integrals are believed to be described by multiple polylogarithms \cite{ArkaniHamed:2012uj}.

The discoveries listed above led to a flurry of activity in computing scattering amplitudes with many particles at high loop orders in planar $\cN=4$ SYM, allowing the computation of the six particle remainder function up to five loops in both the MHV and the NMHV configuration \cite{Caron-Huot:2016owq}. For seven particles the MHV amplitude is known analytically at two loops \cite{Golden:2014xqf} and there are symbol level results at four loops in the MHV case and at three loops in the NMHV case  \cite{Dixon:2016nkn}. Beyond seven particles the associated cluster algebra becomes infinite and the possibility appears of moving outside of the realm of polylogarithmic functions, rendering it difficult to use the techniques utilized to derive the results at lower multiplicities. One way to analyse this bottleneck, is to consider special kinematic limits such as the multi-Regge limit.

In the Regge limit $s \gg |t|$, amplitudes develop large logarithms as was originally studied from the point of view of parton-parton scattering in QCD. The resummation of these logarithms at leading logarithmic (LLA) \cite{Kuraev:1976ge,Kuraev:1977fs,Balitsky:1978ic} and next-to-leading logarithmic accuracy (NLLA) \cite{Fadin:1998py,Camici:1997ij,Ciafaloni:1998gs} was given in the seminal work of Balitsky, Fadin, Kuraev and Lipatov (BFKL). In planar $\cN=4$ SYM, in the Euclidian region, scattering amplitudes are determined to all orders by four- and five-point amplitudes, but exhibit additional contributions, starting from six points, when analytically continued to other Mandelstam regions. They appear as a result of crossing the so called Regge cut \cite{Bartels:2008ce,Bartels:2008sc} and can be described as a dispersion integral which factorizes in Fourier-Mellin space.

From this perspective we continue the line of research from \cite{DelDuca:2016lad} and \cite{DelDuca:2018hrv} in which scattering amplitudes in the multi-Regge limit of $\cN=4$ SYM were considered for any number of particles at LLA and for seven points at NLLA respectively. There, the realisation that in the multi-Regge limit all kinematic dependence is transverse, and thus our configuration space is given by points on the complex plane led to a full understanding of the function space in multi-Regge kinematics (MRK), given by single-valued multiple polylogarithms. This understanding allows one to compute the coefficients appearing in the perturbative expansion of the dispersion integral recursively in loop order via Fourier-Mellin convolutions. Furthermore, the Fourier-Mellin picture allows one to describe certain of these perturbative coefficients for $N$ particle scattering at LLA as linear combinations of perturbative coefficients for scattering amplitudes with lower multiplicity. This constituted an extension of the two-loop factorisation observed in \cite{Bartels:2011ge}. These developments allowed for the computation of the MHV amplitude at LLA for any number of particles to five loops, the amplitude for eight or less particles for any helicity configuration at LLA up to four loops \cite{DelDuca:2016lad} and the seven-point amplitude at NLLA for MHV up to five and up to 3 and 4 loops for the two independent NMHV helicity configurations, respectively \cite{DelDuca:2018hrv}. In this paper, we extend this analysis to the case of eight external particles at NLLA. In particular, we start from the two-loop eight-point remainder function at NLLA and lift it to three loops using Fourier-Mellin convolutions. Afterwards, we use the same framework to compute all non-MHV contributions to the eight-point amplitude and analyse its analytic form and leading singularity structure.

This paper is organised as follows. In Section \ref{sec:intro} we review kinematics in the multi-Regge limit and the function space which it implies. In Section \ref{sec:dispint} we turn to a conjectural representation of the eight-particle amplitude given as a dispersion integral and discuss the consistency of this integral and its interpretation as a Fourier-Mellin transform in the weak coupling limit. In Section \ref{sec:eightpoints}, we study the perturbative expansion of this Fourier-Mellin transform and review the convolution formalism, before applying it to the three-loop amplitude for all helicity configurations and discussing its analytic form. Finally Section \ref{sec:conclu} contains our conclusions and an outlook to future work for generic number of particles. All the results obtained in this paper are provided as ancillary material with the arXiv submission of this paper.

\section{The Multi-Regge Limit of $\mathcal{N} = 4$ SYM}\label{sec:intro}

\subsection{Multi-Regge Kinematics}
In this paper we will consider the scattering of eight gluons in $\cN = 4$ SYM in a special kinematic regime called multi-Regge kinematics (MRK).
To define this limit, we will work with a generic number of particles and consider $1 \, + \, 2 \rightarrow 3 \, + \, \dots + N$ scattering with all particles outgoing. We will work in lightcone coordinates
\begin{align}
&p^\pm \equiv p^0 \pm p^z, & \bp_{k} \equiv p_{k\perp} = p_k^x \pm i p_k^y \,,
\end{align}
with the scalar product of two vectors $p$ and $q$ given by
\begin{align}
2 \, p \cdot q = p^+q^- + p^-q^+ - \bp \bar{\textbf q} - \bar{\bp} {\textbf q} \,.
\end{align}
Without loss of generality, we will choose the reference frame in which the momenta of the initial-state gluons lie on the $z$-axis with $p_2^0 = p_2^z$, implying $p_1^+ = p_2^- = \bp_1 = \bp_2 = 0$.
In this case the multi-Regge limit corresponds to the limit where
\begin{align}
&p_3^+ \gg p_4^+ \gg \dots \gg p_{N-1}^+ \gg p_N^+, & |\bp_3| \simeq \dots \simeq |\bp_N| \,.
\end{align}
The on-shell conditions $p_i^2 = p_i^+ p_i^- - |\bp_i|^2 = 0$ then imply
\begin{align}
&p_3^- \ll p_4^- \ll \dots \ll p_{N-1}^- \ll p_N^- \,.
\end{align}
In addition to the momenta $p_i$, we introduce dual coordinates $x_i$ as
\begin{equation}
x_i - x_{i-1} = p_i \,.
\end{equation}
Amplitudes in planar $\cN = 4$ SYM obey dual conformal invariance, which implies that the kinematical dependence can be expressed in terms of conformal cross-ratios
\beq\label{eq:Ucrossratios}
U_{ij}\equiv\frac{x^2_{i+1j}x^2_{ij+1}}{x^2_{ij}x^2_{i+1j+1}}\,,
\eeq
where $x_{ij} = x_i - x_j$ and indices are cyclically identified, $i+N\simeq i$. Of these, only $3N-15$ are algebraically independent in four dimensions.

In the multi-Regge-limit, amplitudes only depend on the transverse momenta $\bp_i$ or equivalently the  \textit{transverse dual coordinates} $\bx_i$ defined by
\beq
\bp_{i+3} \equiv \bx_{i+2} - \bx_{i+1}, \quad 0 \leq i \leq N-4,
\eeq
Since the kinematics is determined entirely by the $\bx_i$ which obey dual conformal symmetry, we can identify the configuration space as $N-2$ points in $\mathbb{CP}^1$, or equivalently the moduli space of genus zero curves with $N-2$ marked points $\mathfrak{M}_{0,N-2}$.
Upon fixing the residual $SL(2,\mathbb{C})$ symmetry of $\mathfrak{M}_{0,N-2}$, we can find various coordinate systems corresponding to $N-5$ cross ratios formed out of the dual coordinates $\bx_i$, such as
\begin{equation}\label{eq:z_i_def}
z_i\equiv  \frac{({\textbf x}_1 -{\textbf x}_{i+3})\,({\textbf x}_{i+2} -{\textbf x}_{i+1})}{({\textbf x}_1 -{\textbf x}_{i+1})\,({\textbf x}_{i+2} -{\textbf x}_{i+3})}  \,.
\end{equation}
These coordinates are well-suited to describe the Fourier-Mellin transforms in amplitudes in MRK, and we will hence refer to them as \emph{Fourier-Mellin coordinates}.  A set of local coordinates in which the functional dependence of MRK amplitudes is particularly simple are the \emph{simplicial coordinates}, obtained by fixing three of the $N-2$ points to $0$, $1$, and $\infty$.
A special set of simplicial coordinates named \emph{simplicial MRK coordinates} is given by
\begin{equation}
(\bx_1, \dots, \bx_{N-2}) \rightarrow (1,0, \rho_1, \dots, \rho_{N-5}, \infty)\, .
\end{equation}
They are related to the Fourier-Mellin coordinates by
\begin{equation}\label{ztorho}
z_i=\frac{(1-\rho_{i+1})(\rho_i-\rho_{i-1})}{(1-\rho_{i-1})(\rho_i-\rho_{i+1})}\,,
\quad
\rho_0=0\,,\rho_{N-4}=\infty\,.
\end{equation}
This set of coordinates is particularly interesting, as it was shown first for the two loop symbol to NLLA in \cite{Bargheer:2015djt} and later to all orders at LLA at function level in \cite{DelDuca:2016lad} that multi-leg amplitudes in the multi-Regge limit of $\cN = 4$ SYM factorize into a finite number of building blocks with fewer legs when expressed in these coordinates. This property was used to compute the scattering of $N$ particles up to five loops at LLA for MHV amplitudes.

\subsection{Single-Valued Polylogarithms}\label{sec:polys}

As a consequence of the kinematics being described entirely by $\mathfrak{M}_{0,N-2}$, amplitudes in MRK correspond to a family of iterated integrals called multiple polylogarithms (MPLs) \cite{Goncharov:2001}, defined by the recursion
\beq
G(a_1,\ldots,a_n; z) = \int_0^z\frac{dt}{t-a_1}G(a_2,\ldots,a_n;t)\,,
\eeq
with $G(;z)=1$. The number of integrations $n$ is called the weight. In the case where all the $a_i$ are zero, we define
\beq
G(\underbrace{0,\ldots,0}_{n\textrm{ times}}\,;z)= \frac{1}{n!}\log^nz\,.
\eeq
In the multi-Regge-limit, the function space can be restricted even further. It was shown already in \cite{Dixon:2012yy,DelDuca:2016lad} that amplitudes at LLA in the multi-Regge limit are single-valued, an analysis which will be extended to all orders in \cite{FutureCite}. MPLs, however, generally exhibit branch cuts and it would be preferable to write the amplitude in terms of manifestly single-valued objects.

For this purpose, we will consider linear combinations of multiple polylogarithms such that their branch cuts cancel. More specifically, we can associate to every multiple polylogarithm\footnote{In what follows, we use the shorthand $G_{a_1,\dots,a_n}(z) \equiv G(a_1,\dots,a_n;z)$.} $G_{a_1,\dots,a_n}(z)$ a so called a single-valued multiple polylogarithm (SVMPL) $\cG_{a_1,\dots,a_n}(z)$. The SVMPL is a linear combination of multiple polylogarithms in the variable $z$ and its complex conjugate $\zb$ such that it is single-valued and it obeys the same holomorphic differential equation as the original multiple polylogarithm
\beq
\partial_z \cG_{a_1,\dots,a_n}(z) = \frac{1}{z-a_1}\cG_{a_2,\dots,a_n}(z).
\eeq
One can show \cite{BrownSVMPLs,Schnetz:2013tu} that there is a one-to-one map which sends each MPL to its single-valued analogue
\beq
{\textbf{s}}\left(G_{a_1,\dots,a_n}(z)\right)\equiv\cG_{a_1,\dots,a_n}(z).
\eeq
For an explicit construction and detailed discussion of this map, see \cite{DelDuca:2016lad}. Its action on weight one and two multiple polylogarithms is given by
\begin{align}
{\textbf{s}}\left(G_{a}(z)\right)=\cG_{a}(z)&=G_{a}(z)+G_{\bar a}(\bar z),\label{eq:ExampleSVMPL}\\
{\textbf{s}}\left(G_{a,b}(z)\right)=\cG_{a,b}(z)&= G_{a,b}(z)+G_{\bar{b},\bar{a}}(\bar{z})+G_{b}(a) G_{\bar{a}}(\bar{z})+G_{\bar{b}}(\bar{a}) G_{\bar{a}}(\bar{z})\nonumber\\
&-G_{a}(b) G_{\bar{b}}(\bar{z})+G_{a}(z) G_{\bar{b}}(\bar{z})-G_{\bar{a}}(\bar{b}) G_{\bar{b}}(\bar{z})\,.
\end{align}
As an example, let us have a closer look at the weight one case
\beq
\cG_a(z) = G_a(z)+G_{\bar{a}}(\zb)= \log\left(1-\frac{z}{a}\right) +\log\left(1-\frac{\zb}{\bar{a}}\right)=\log \left| 1-\frac{z}{a} \right|^2.
\eeq
Since the argument of the logarithm on the right-hand side is positive-definite, we see explicitly that the function is single-valued. Single-valued multiple polylogarithms inherit many of the properties of ordinary MPLs. In particular, SVMPLs form a shuffle algebra and satisfy many of the same functional relations as their multi-valued analogues.

\section{Amplitudes in the Multi-Regge Limit of $\mathcal{N}=4$ SYM}\label{sec:dispint}

\subsection{The MRK Ratio Function}
Let us now turn to the representation of amplitudes in MRK in $\cN = 4$ SYM.
Helicity must be conserved among the gluons going very forward, so that we only distinguish between different helicity configurations $(h_1, \dots h_{N-4})$ of the gluons emitted along the ladder.
Let us define the ratio
\begin{equation}\label{eq:R_definition}
e^{i\Phi_{h_1,\ldots,h_{N-4}}}\,\cR_{h_1\ldots h_{N-4}} \equiv \left[\frac{A_N(-,+,h_1,\ldots,h_{N-4},+,-)}{A_N^{\textrm{BDS}}(-,+,\ldots,+,-)}\right]_{|\textrm{MRK}}\,,
\end{equation}
where $A_N(-,+,h_1,\ldots,h_{N-4},+,-)$ is the (colour-ordered) amplitude for the production of $N-4$ gluons emitted along the ladder, and $A_N^{\textrm{BDS}}(-,+,\ldots,+,-)$ is the corresponding BDS amplitude.

The term $\cR_{h_1 \ldots h_{N-4}}$ is finite and dual conformally invariant and is related to the well-known remainder and ratio functions.
In the Euclidean region, the ratio tends to a phase, which is immaterial for further considerations in this paper and we will hence normalize the left hand side such that $\cR_{h_1 \ldots h_{N-4}} = 1$ in the Euclidean region.
Performing an analytic continuation in the final state momenta to a different \emph{Mandelstam region}, $\cR_{h_1 \ldots h_{N-4}}$ will no longer be trivial due to the presence of a Regge cut \cite{Bartels:2014mka,Bartels:2008ce,Bartels:2008sc,Lipatov:2010ad,Lipatov:2010qg,Bartels:2011ge,Lipatov:2012gk,Bartels:2013jna,Bartels:2014jya}.

\subsection{The $8$ point Dispersion Integral at Weak Coupling}

\begin{figure}[H]
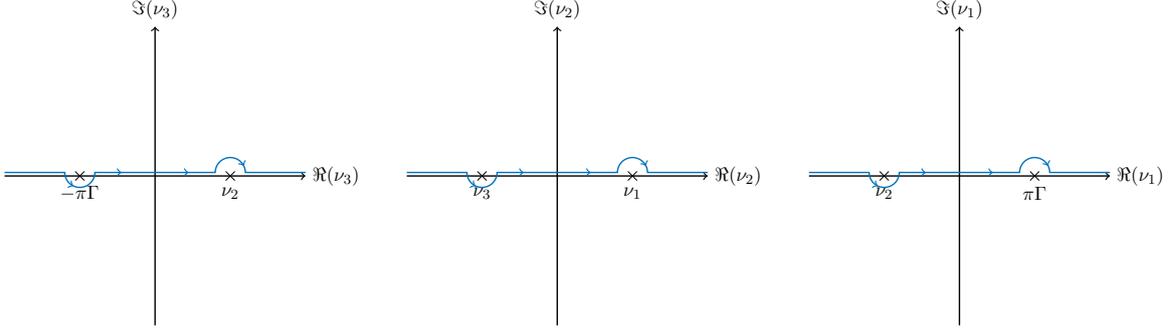

        \centering
        \begin{minipage}{0.3\textwidth}
        \centering
        \scalebox{0.66}{\contourOne}
        \end{minipage}\hfill
        \begin{minipage}{0.3\textwidth}
        \centering
        \scalebox{0.66}{\contourTwo}
        \end{minipage}\hfill
        \begin{minipage}{0.3\textwidth}
        \centering
        \scalebox{0.66}{\contourThree}
\end{minipage}\caption{The initial integration contour for the $8$-gluon BFKL integral.}
        \label{fig:contourRNInitial}
\end{figure}

In the following sections, we will consider the case of $N = 8$ particles.
In \cite{FutureCite}, inspired by work done at LLA \cite{Lipatov:2010ad, Bartels:2011ge, DelDuca:2016lad}, it will be shown that after analytically continuing the momenta $p_4 $ to $ p_7$ to negative energies, $\cR_{h_1h_2h_3h_4}$ is given by the relation
\begin{equation}
\begin{split}\label{eq:MRK_conjecture}
\cR_{h_1h_2h_3h_4}e^{i \delta_8}= &1  + a\,i\pi\, \cF_3 \left[\left(\prod_{k=1}^{3}e^{-L_k \omega_k}\right)\, \chi^{h_1}_{1} C^{h_{2}}_{1,2}C^{h_{3}}_{2,3} \,\chi^{-h_{4}}_{3} \right] \,,
\end{split}
\end{equation}where $\cF_{m} $ denotes the $m$-fold inverse Fourier-Mellin transform,
\begin{equation}
\cF_{m}\bigg[f(\lbrace \nu_i,n_i \rbrace)\bigg] \equiv \prod_{k=1}^{m}\sum_{n_k=-\infty}^{+\infty}\left(\frac{z_k}{\zbar_k}\right)^{n_k/2}\int_{-\infty}^{+\infty}\frac{d\nu_k}{2\pi}|z_k|^{2i\nu_k} f(\lbrace \nu_i,n_i \rbrace) \,,
\end{equation}
whose contour of integration is shown in fig. \ref{fig:contourRNInitial}.
The dependent variables in \eqref{eq:MRK_conjecture} are implicit on the left hand side of the equation, $a$ is the 't Hooft coupling and  $L_k = \log \tau_k + i \pi$, where $\log \tau_k$ are the large logarithms with
 \beq
 \tau_k \equiv \sqrt{U_{8,k+2}\,U_{1,k+3}}\, \xrightarrow[\text{MRK}]{ }\, 0.
 \eeq
The integrand factorises into building blocks
\begin{align}
  \omega(\nu_i,n_i) &\equiv \omega_i = -a(E_i + a E_i^{(1)} + \cO(a^2)),\\
  \chi^{\pm}(\nu_i,n_i)&\equiv \chi_i^{\pm} = \chi_{0,i}^{\pm} (1 + a \kappa_1^\pm + \cO(a^2)) ,\\
  C^{\pm}(\nu_i,n_i,\nu_j,n_j) &\equiv C^{\pm}_{i,j} = C^{\pm}_{0,i,j} (1 + a c_{1,i,j}^\pm + \cO(a^2))  ,
\end{align}
which are the all-order BFKL eigenvalue, the all order impact factor and the all order central emission block respectively. We shall collectively refer to these objects as the BFKL building blocks. The exact form of the BFKL building blocks up to NLO can be found in appendix \ref{app:BBB}.
\begin{figure}[h!]
  \centering
  \includegraphics[width=0.4\textwidth]{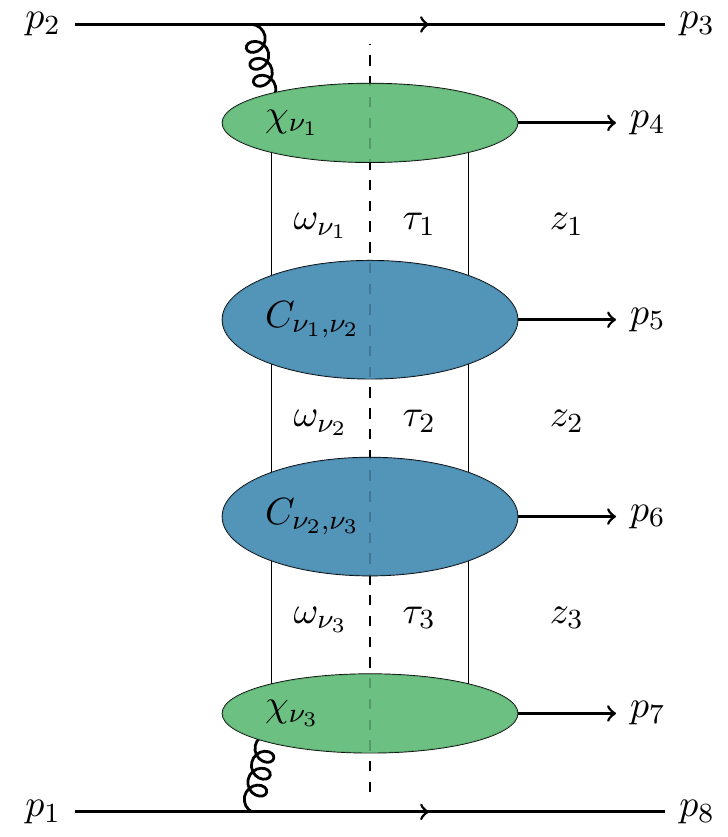}
  \caption{The structure of the $8$ point amplitude in MRK.}
\end{figure}

The BDS phase $\delta_8$ is given by
\begin{equation}\label{eq:BDS_Phase}
\delta_8 = \pi \Gamma \log \left| \frac{\bx_{32}\bx_{61} \bx_{56}\bx_{21}}{\bx_{31}\bx_{62} \bx_{51}\bx_{26}} \right|^2
= \pi \Gamma \log \left|\frac{\rho_{1}}{(\rho_1 - 1)(\rho_{3} - 1)} \right|^2,
\end{equation}
where $\Gamma$ is proportional to the cusp anomalous dimension $\Gamma \equiv \frac{\gamma_K}{8}=\frac{a}{2}-\frac{\zeta_2 a^2}{2}+\mathcal{O}(a^3)$ known to all orders from integrability.
As will be shown in \cite{FutureCite} the form of the contours in \eqref{eq:MRK_conjecture} is dictated by the correct behaviour under soft limits, as was done for the weak coupling expansion for seven external particles in \cite{DelDuca:2018hrv}. In the limit where a gluon becomes soft, we have
\begin{align}
\lim_{z_1 \rightarrow 0} \cR_8 e^{i \delta_8} &= |z_1|^{2 \pi i \Gamma} \left(\cR_7e^{ i \delta_{7}}\right)(z_2,z_3)  \, ,  \label{eq:soft1}\\
\lim_{z_2 \rightarrow 0, z_{1}z_{2}\text{ fixed}}  \cR_8e^{i \delta_8} &=  \left(\cR_7 e^{i \delta_{7}} \right) (-z_1z_2 ,z_3)\label{eq:soft2} \, ,\\
\lim_{z_3 \rightarrow 0, z_{2}z_{3}\text{ fixed}} \cR_8 e^{i \delta_8} &= \left(\cR_7 e^{i \delta_{7}} \right) (z_1,-z_{2}z_3) \label{eq:soft3}  \, ,\\
\lim_{z_{3} \rightarrow \infty} \cR_8 e^{i \delta_8} &=|z_3|^{-2 \pi i \Gamma} \left( \cR_7 e^{i \delta_{7}} \right)(z_1,z_{2})  \, , \label{eq:soft4}
\end{align}
we can infer that for $i \in \{2,3\}$, the right hand side has poles at $\nu_1 = \pi \Gamma - i 0^+$, $\nu_i = \nu_{i-1} + i 0^+$, and $\nu_{3} = - \pi \Gamma + i 0^+$ for $n_1 = 0$, $n_{i-1} = n_i$, and $n_{3} = 0$ respectively. This follows recursively starting from the six- and seven-point cases analysed in \cite{DelDuca:2018hrv}.

Analogously to the seven-point case, the integral would develop a pinch singularity in the weak coupling limit when $n_{i-1} = n_{i} = 0$ for $i \in \{2,3\}$ whenever there is no insertion of a BFKL eigenvalue in the integrand. This singularity can be regularized by deforming the integration contour. In particular, this may be done by subtracting either the residues at $\nu_1 = \pi \Gamma$ and at $\nu_2 = \nu_3$ or at $\nu_1 = \nu_2$ and at $\nu_{3} = -\pi \Gamma$.
\noindent To facilitate the deformation of the contour we will rewrite our integrand in terms of an alternative set of building blocks, namely
\begin{align}
\tilde{C}^h(\nu_1,n_1,\nu_2,n_2) &= \frac{C^h(\nu_1,n_1,\nu_2,n_2)}{\chi^-(\nu_1,n1) \chi^+(\nu_2,n_2)}, \\
\tilde{\Phi}(\nu,n) &= \chi^+(\nu,n) \chi^-(\nu,n), \\
I^h(\nu,n) &\equiv \frac{\chi^h(\nu,n)}{\chi^+(\nu,n)} = \begin{cases} 1, &h=+ \\ H(\nu,n), &h=- \end{cases} \quad ,
\end{align}
where
\begin{equation}
H(\nu,n) = \frac{\chi^-(\nu,n)}{\chi^+(\nu,n)},
\end{equation}
is the helicity flip kernel in Fourier-Mellin space known to all orders from \cite{Basso:2014pla}. Expressed in these building blocks the combination of impact factors and central emission blocks reads
\begin{align}
\chi^{h_1}_{1} C^{h_{2}}_{1,2}C^{h_{3}}_{2,3} \,\chi^{-h_{4}}_{3} = I_1^{h_1}\tilde{\Phi}_{1}\tilde{C}^{h_{2}}_{1,2} \tilde{\Phi}_{2}\tilde{C}^{h_{3}}_{2,3} \tilde{\Phi}_{3} \,\bar{I}^{h_{4}}_{3}.
\end{align}
The poles at $\nu_2 = \pm \pi \Gamma$  and the poles at $\nu_1 = -\pi \Gamma$ and at $\nu_{3} = \pi \Gamma$ that are seemingly introduced by the $\tilde{\Phi}$ in between the pairs of $C$'s are spurious, as the central emission block vanishes for these values  \cite{DelDuca:2018hrv}
\begin{equation}
C^h(- \pi \Gamma, 0 , \nu, n) = C^h(\nu,n,\pi \Gamma, 0) = 0.
\end{equation}
In \cite{DelDuca:2018hrv}, the behaviour of the alternative building blocks was determined to be
\begin{align}
\omega(\pm \pi \Gamma, 0) &= 0 \, , \\
I(\nu,0) &= 1 \, , \\
\tilde{C}^h(\pi \Gamma, 0 , \nu, n) &= i \pi a I^h(\nu,n)  \, ,\\
\tilde{C}^h( \nu, n,-\pi \Gamma, 0 ) &= -i \pi a \bar{I}^h(\nu,n) \, , \\
\Res_{\nu_1= \nu_2}\tilde{C}^h(\nu1,n,\nu_2,n) &= \frac{(-1)^{n+1} i e^{i \pi \omega(\nu_2,n)}}{\tilde{\Phi}(\nu_2,n)}  \, ,\\
\Res_{\nu = \pm \pi \Gamma } \tilde{\Phi}(\nu,0) &= \pm \frac{1}{\pi a}.
\end{align}

\begin{figure}[h]
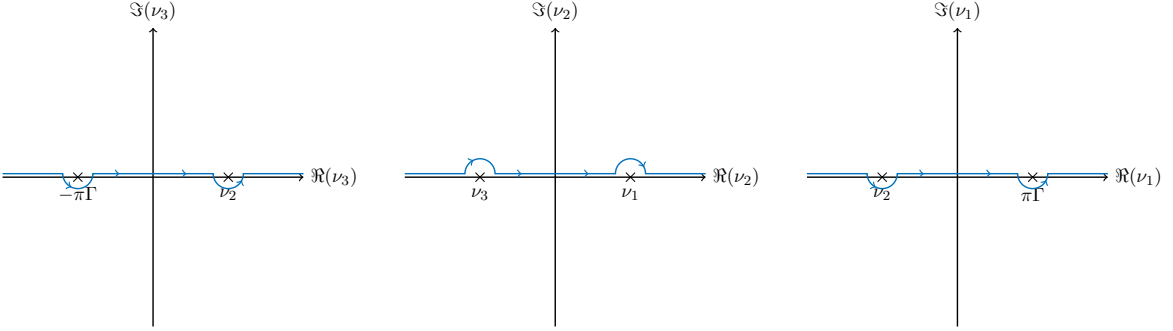

        \centering
        \begin{minipage}{0.3\textwidth}
        \centering
        \scalebox{0.66}{\contourOneShifted}
        \end{minipage}\hfill
        \begin{minipage}{0.3\textwidth}
        \centering
        \scalebox{0.66}{\contourTwoShifted}
        \end{minipage}\hfill
        \begin{minipage}{0.3\textwidth}
        \centering
        \scalebox{0.66}{\contourThreeShifted}
\end{minipage}\caption{The integration contour for the regularized version of the $8$-gluon BFKL integral.}
        \label{fig:contourRNFinal}
\end{figure}

With this at hand we can now explicitly perform the necessary contour deformations by computing the corresponding residues. Starting with the residue at $\nu_1 = \pi \Gamma$, we find that upon subtracting the residues at $\nu_1 = \pi \Gamma$ and at $\nu_2 = \nu_3$, we get
\begin{align}\label{eq:contourFixN}
\begin{split}
\cR_{h_1h_2h_3h_4} e^{i \delta_8} = \, &|z_1|^{2 \pi i \Gamma} \left(\cR_{h_2h_3h_4} e^{i \delta_{7}}\right)(z_2,z_3) +  \left(\cR_{h_1h_2h_4} e^{i \delta_{7}}\right)(z_1,-z_2z_3)  \\
&- \left(\cR_{h_2h_4} e^{i \delta_6}\right)(-z_2z_3) + 2 \pi i f_{h_1h_2h_3h_4},
\end{split}
\end{align}
where
\begin{equation}\label{eq:def_f}
\begin{split}
f_{h_1h_2h_3h_4} =\; &\frac{a}{2}\,i\pi \left[ \prod_{k=1}^{3}\sum_{n_k=-\infty}^{+\infty}\left(\frac{z_k}{\zbar_k}\right)^{\frac{n_k}{2}}\int_{-\infty}^{+\infty}\frac{d\nu_k}{2\pi}|z_k|^{2i\nu_k} \right] \\
&\times\left[\prod_{k=1}^{3}e^{-L_k \omega_k}\right]\, I_1^{h_1}\tilde{\Phi}_{1}\tilde{C}^{h_{2}}_{1,2} \tilde{\Phi}_{2}\tilde{C}^{h_{3}}_{2,3} \tilde{\Phi}_{3}    \,\bar{I}^{h_{4}}_{3},
\end{split}
\end{equation}
and where the contours of integration correspond to the ones depicted in fig. \ref{fig:contourRNFinal}.

As the integral \eqref{eq:def_f} exhibits no pinch singularities, we can use the relation \eqref{eq:contourFixN} to define the regularized Fourier-Mellin integral. At two loops NLLA we define
\begin{align}\label{eq:def_regul}
\cF_3 \left[ \cI_{\text{NLLA}}^{(2)} \right]  &= \cF_3^{\text{Reg}} \left[ \cI_{\text{NLLA}}^{(2)} \right] \\
& =  \prod_{k=1}^{3}\sum_{n_k=-\infty}^{+\infty}\left(\frac{z_k}{\zbar_k}\right)^{\frac{n_k}{2}}\int_{-\infty}^{+\infty}\frac{d\nu_k}{2\pi}|z_k|^{2i\nu_k} \cI_{\text{NLLA}}^{(2)}  + P^{(2;0,0,0)}_{h_1h_2h_3h_4} + 2 \pi i \,Q^{(2;0,0,0)}_{h_1h_2h_3h_4} \,, \nonumber
\end{align}
where the contours of integration on the r.h.s. correspond to the ones depicted in fig. \ref{fig:contourRNFinal}, $\cI_{\text{NLLA}}^{(2)} $ is the 2 loop NLLA contribution to the integrand
\begin{equation}
\cI = \left(\prod_{k=1}^{3}e^{-L_k \omega_k}\right)\, \chi^{h_1}_{1} C^{h_{2}}_{1,2}C^{h_{3}}_{2,3} \,\chi^{-h_{4}}_{3},
\end{equation}
of \eqref{eq:MRK_conjecture} and where $P^{(2;0,0,0)}_{h_1h_2h_3h_4}$ and $Q^{(2;0,0,0)}_{h_1h_2h_3h_4}$ are the two-loop contributions of the lower-point amplitudes on the right-hand side of \eqref{eq:contourFixN}, an explicit form of which will be given in \eqref{eq:def_P} and \eqref{eq:def_Q}.

Having taken care of the two-loop case, we now have a consistent Fourier-Mellin representation for the eight particle ratio function in MRK to all orders at NLLA. In what follows, we will use the properties of the Fourier-Mellin integral to compute the perturbative expansion of this object.

\section{The Eight-Point Amplitude at NLLA in MRK}\label{sec:eightpoints}

\subsection{A Fourier-Mellin Representation for the Eight-Point Amplitude in MRK}

As our purpose is to compute the eight-point amplitude in the multi-Regge limit to a certain loop order, we are interested in the perturbative expansion of $\cR_{h_1 h_2 h_3 h_4}$. To this end, we define
\begin{equation}\label{eq:perturbativeExpansion}
\begin{split}
\cR_{h_1 h_2 h_3 h_{4}}&\left(\tau_{1},{z_{1}},\tau_{2},{z_{2}},\tau_{3},{z_{3}}\right)e^{i \delta_8}  =1+ 2\pi i \sum_{\ell=1}^{\infty}\sum_{i_1,i_2,i_3=0}^{\ell-1}a^\ell \,\left(\prod_{k=1}^{3}\frac{1}{i_k!}\log^{i_k}\tau_k\right) \\
&\times \left( \,{\tilde g}_{h_1h_2h_3h_4}^{(\ell;i_1,i_2,i_{3})}(z_1,z_2,z_{3}) + 2 \pi i \, {\tilde h}_{h_1h_2h_3h_4}^{(\ell;i_1,i_2,i_3)}(z_1,z_2,z_{3}) \right)\,.
\end{split}
\end{equation}
In the following, the main focus will lie on the \emph{perturbative coefficients} ${\tilde g}_{h_1h_2h_3h_4}^{(\ell;i_1,i_2,i_{3})}$ and ${\tilde h}_{h_1h_2h_3h_4}^{(\ell;i_1,i_2,i_{3})}$ up to NLLA, i.e. $\ell - 2 \leq \sum_{k=1}^{N-5} i_k \leq \ell - 1$, and we will drop explicit dependence on the Fourier Mellin coordinates $\{z_i \}$. Note that the LLA remainder function is purely imaginary and hence
\begin{equation}
{\tilde h}_{h_1h_2h_3h_4}^{(i_1+i_2+i_3+1;i_1,i_2,i_3)} = 0\,.
\end{equation}
The ratio $\cR_{h_1 h_2 h_3 h_{4}}$ is proportional to an inverse Fourier-Mellin transform and thus, by expanding the right hand side of eq. \eqref{eq:MRK_conjecture}, the perturbative coefficients themselves can also be associated to inverse Fourier-Mellin transforms.
For simplicity, we define a shorthand for the product of leading-order impact factors and central emission blocks, which we will refer to as the \emph{vacuum ladder}
\begin{equation}
\varpi_{8} \equiv \chi^{h_1}_{0,1} C^{h_{2}}_{0,1,2} C^{h_{3}}_{0,2,3}\chi^{-h_{4}}_{0,3} \,,
\end{equation}
where we drop explicit dependence on the helicities.
Then, at LLA (i.e. $\sum_{k=1}^{3}i_k = \ell - 1$) we find
\begin{align}
{\tilde g}_{h_1h_2h_3h_4}^{(\ell;i_1,i_2,i_{3})} =
-\frac{1}{2} \,\cF_{3} \Bigg[ \bigg(\prod_{k=1}^{3} E^{i_k}_{\nu_k,n_k}\bigg) \varpi_{8} \Bigg]\,.
\end{align}
At NLLA, (i.e. for  $\sum_{k=1}^{3}i_k = \ell - 2$), we write
\begin{equation}
\begin{split}
\label{eq:inrto_corrected_PertCoef}
{\tilde g}_{h_1h_2h_3h_4}^{(\ell;i_1,i_2,i_{3})}&= \sum_{j=1}^{3} i_j {\tilde{g}}_{h_1h_2h_3h_4}^{j;(\ell;i_1,i_2,i_{3})} +  {\tilde{g}}_{\varpi;h_1h_2h_3h_4}^{(\ell;i_1,i_2,i_{3})}\,,\\
{\tilde h}_{h_1h_2h_3h_4}^{(\ell;i_1,i_2,i_{3})} &= \sum_{j=1}^{3}  {\tilde{h}}_{h_1h_2h_3h_4}^{j;(\ell;i_1,i_2,i_{3})} +  {\tilde{h}}_{\varpi;h_1h_2h_3h_4}^{(\ell;i_1,i_2,i_{3})}\, ,
\end{split}
\end{equation}
 where the perturbative coefficients with extra indices are called \emph{corrected perturbative coefficients}. Perturbative coefficients with an additional upper index correspond to insertions of the NLO corrections to the BFKL eigenvalue and perturbative coefficients with an additional lower index correspond to the insertion of the corrections to the vacuum ladder.
 This implies
  \begin{equation}
  \begin{split}\label{eq:g_NLLA_def}
{\tilde{g}}_{h_1h_2h_3h_4}^{j;(\ell;i_1,i_2,i_{3})} =& \,
  \frac{1}{2} \,\cF_{3} \left[\varpi_{8} \, E^{(1)}_j\, \prod_{k=1}^{3} E^{i_k-\delta_{kj}}_k\right],\\
{\tilde{g}}_{\varpi;h_1h_2h_3h_4}^{(\ell;i_1,i_2,i_{3})} =& \,
  \frac{1}{2} \,\cF_{3} \left[\varpi_{8} \, \left(\kappa^{h_1}_{1,1}+ \Re \left( c^{h_2}_{1,1,2}\right)+\Re \left( c^{h_3}_{1,2,3}\right) +\kappa^{-h_{4}}_{1,3}\right) \, \prod_{k=1}^{3} E^{i_k}_k\right].
  \end{split}
  \end{equation}
for the imaginary parts and
 \begin{equation}
 \begin{split}\label{eq:h_NLLA_def}
{\tilde{h}}_{h_1h_2h_3h_4}^{j;(\ell;i_1,i_2,i_{3})} =& \, -\frac{1}{4} \,\cF_{3} \left[\varpi_{8} \, E_j\, \prod_{k=1}^{3} E^{i_k}_k\right],\\
{\tilde{h}}_{\varpi;h_1h_2h_3h_4}^{(\ell;i_1,i_2,i_{3})} =& \, \frac{1}{4\pi} \,\cF_{3} \left[\varpi_{8} \, \left(\Im \left( c^{h_2}_{1,1,2}\right)+\Im \left( c^{h_3}_{1,2,3}\right)\right)\, \prod_{k=1}^{3} E^{i_k}_k\right],
\end{split}
 \end{equation}
for the real contributions, where $\Re$ and $\Im$ denote the real- and imaginary parts, respectively. In what follows, we will elaborate on a method to compute the corrected perturbative coefficients and determine them for the three loop case.

\subsection{Fourier-Mellin Convolutions for Amplitudes in MRK}\label{subsec:convo}

As we saw in the previous section, the perturbative coefficients correspond to a three-fold inverse Fourier-Mellin transform.
This transformation maps products into convolutions, so that for $\cF[F]=f$ and $\cF[G]=g$ we have
\begin{align}\label{eq:conv_thm}
\cF[F\cdot G] = \cF[F]\ast\cF[G] = f\ast g\,,
\end{align}
where the convolution is given by
\begin{align}\label{eq:conv_def}
(f\ast g)(z) = \frac{1}{\pi}\int \frac{d^2w}{|w|^2}\,f(w)\,\,g\left(\frac{z}{w}\right)\,.
\end{align}
 Using this convolution product for Fourier-Mellin transforms, we can identify simple relations between the perturbative coefficients at different loop orders. At six points and at LLA, for example, we have
\begin{align}
g_{++}^{(\ell;\ell-1)} = -\frac{1}{2}\cF \left[ \chi^+ E^{\ell-1} \chi^- \right] = g_{++}^{(\ell-1;\ell-2)} * \cF[E] \,.
\end{align}
By repeatedly convoluting with leading-order BFKL eigenvalues, we can compute higher-loop perturbative coefficients recursively from known lower-loop results.
Since the perturbative coefficients are single-valued \cite{DelDuca:2016lad}, the evaluation of the convolution integrals can be simplified to a residue computation, as was shown in \cite{Schnetz:2013hqa}. Let $f(z)$ be a linear combination of single-valued polylogarithms over rational functions with singularities at $z=a_i$ and $z = \infty$. Close to each singularity, $f$ can be written as
\begin{align}
f(z) &\,= \sum_{k,m,n}\,c^{a_i}_{k,m,n}\,\log^k\left|1-\frac{z}{a_i}\right|^2\,(z-a_i)^m\,(\zb-\bar{a}_i)^n\,, \quad z\to a_i\,,\\
f(z) &\,= \sum_{k,m,n}\,c^{\infty}_{k,m,n}\,\log^k\frac{1}{|z|^2}\,\frac{1}{z^m}\,\frac{1}{\zb^n}\,, \quad z\to \infty\,.
\end{align}
Then we define the {holomorphic residue} of $f$ at $z = a$ as the coefficient of the simple holomorphic pole with no logarithmic singularities,
\begin{equation}
\textrm{Res}_{z=a}f(z) \equiv c^{a}_{0,-1,0}\,.
\end{equation}
The integral of $f$ over the whole complex plane, if it exists, is given by the sum of the holomorphic residues of its single-valued antiholomorphic primitive $F$, i.e. if $\bar{\partial} F = f$, then \cite{Schnetz:2013tu}
\begin{equation}
\int \frac{d^2z}{\pi}\,f(z) = \textrm{Res}_{z=\infty}F(z) - \sum_i\text{Res}_{z=a_i}F(z)\,.
\end{equation}
Thus, using the Fourier-Mellin transform of the leading-order BFKL eigenvalue $\cF(E_i)$,
\begin{equation}
\cE(z_i) \equiv \cF(E_i) = -\frac{z_i+\zb_i}{2\,|1-z_i|^2}\,,
\end{equation}
we can obtain higher order perturbative coefficients through convolutions with lower order objects.
The convolution formalism can also be extended to the computation of non-MHV amplitudes \cite{DelDuca:2016lad}. We can flip the helicities of amplitudes by convolution with the {helicity flip kernel}
\begin{equation}\label{eq:HKer}
\begin{split}
\cH(z) &= \cF\left[\frac{\chi^-_i}{\chi^+_i} \right] = \cH^{(0)}(z_i) + a \cH^{(1)}(z_i) + \cO(a^2)\\
&= \, -\frac{z_i}{(1-z_i)^2} + \frac{a}{4} \left(\cG_1(z_i)+\frac{z_i }{(1-z_i)}\cG_0(z_i)+\frac{z_i }{(1-z_i)^2} \cG_{0,0}(z_i)\right) + \cO(a^2) \,.
\end{split}
\end{equation}
When flipping the helicity of an impact factor, for example we find
\begin{align}
\cF\left[\chi^+({\nu,n})\, F(\nu,n)\right] &\longrightarrow \,\cF\left[\chi^-({\nu,n})\, F(\nu,n)\right] \\
&= \, \cF\left[\frac{\chi^-({\nu,n})}{\chi^+({\nu,n})} \right]\ast\cF\left[\chi^+({\nu,n})\,F(\nu,n)\right]\\
\nonumber&= \,  \cH(z)\ast\cF\left[\chi^+({\nu,n})\,F(\nu,n)\right]\,.
\end{align}
The same kernel can also be used to flip the helicity of one of the central emission blocks, as can be inferred by looking at the seven-point amplitude. Since the MHV amplitudes with all helicities positive and all helicities negative are identical, we can achieve a helicity flip on the central emission by flipping the helicities on the outer emissions:
\begin{align}\label{eq:HFlip_C}
g_{+++}^{(\ell;i_1,i_2)} \longrightarrow g_{+-+}^{(\ell;i_1,i_2)} = \cH(\bar{z}_1) \ast \cH(z_2) \ast g_{---}^{(\ell;i_1,i_2)} = \cH(\bar{z}_1) \ast \cH(z_2) \ast g_{+++}^{(\ell;i_1,i_2)}.
\end{align}
In what follows, we will exploit the methods described above beyond LLA and use them to compute the three loop eight particle amplitude at NLLA in planar $\cN = 4$ SYM.

\subsection{The Eight-Point Three-Loop Amplitude at NLLA}

In this section, we lay out the computation of the three loop eight-point amplitude at NLLA in planar $\cN=4$ SYM. For the MHV case, this amounts to computing the perturbative coefficients
\begin{equation}
\begin{split}
{\tilde g}_{++++}^{(3;\delta_{k,1},\delta_{k,2},\delta_{k,3})}&= {\tilde{g}}_{++++}^{k;(3;\delta_{k,1},\delta_{k,2},\delta_{k,3})} +  {\tilde{g}}_{\varpi;++++}^{(3;\delta_{k,1},\delta_{k,2},\delta_{k,3})}\,,\\
{\tilde h}_{++++}^{(3;\delta_{k,1},\delta_{k,2},\delta_{k,3})} &=  {\tilde{h}}_{++++}^{k;(3;\delta_{k,1},\delta_{k,2},\delta_{k,3})} +  {\tilde{h}}_{\varpi;++++}^{(3;\delta_{k,1},\delta_{k,2},\delta_{k,3})}\,.
\end{split}
\end{equation}
It is clear from their definitions \eqref{eq:g_NLLA_def} and \eqref{eq:h_NLLA_def} that the corrected perturbative coefficients at different loop orders are related by convolutions. Using the methods demonstrated at LLA in the previous section, we can compute the terms given above seperately and compose them to give the full three loop amplitude at NLLA.

First note that the corrected perturbative coefficients $\tilde{g}_{h_1 h_2 h_3 h_{4}}^{j;(\ell;i_1,i_2 ,i_{3})}$ appear only starting from three loops, so we cannot obtain them recursively from the two-loop result and must compute them separately from the remaining terms. At three loops NLLA we can only have a single $i_k = 1$ while all others are equal to zero. Having only a single BFKL eigenvalue in the integrand, the contributions
\begin{equation}
\tilde{g}_{++++}^{k;(3;\delta_{k,1},\delta_{k,2},\delta_{k,3})}(\rho_1,\rho_2,\rho_3) = \tilde{g}_{++}^{1;(3;1)}(\rho_k),
\end{equation}
correspond to six-point corrected perturbative coefficients using a similar argument to the factorization at LLA in \cite{DelDuca:2016lad}. The terms ${\tilde{h}}_{++++}^{k;(3;\delta_{k,1},\delta_{k,2},\delta_{k,3})}$ can be associated to perturbative coefficients at leading logarithmic accuracy found in \cite{DelDuca:2016lad}.

The remaining corrected perturbative coefficients can be obtained via the methods described in the previous section. Our starting points are the corrected perturbative coefficients at two loops, $g_{\varpi;++++}^{(2;0,0,0)}$ and $h_{\varpi;++++}^{(2;0,0,0)}$ which can be obtained from the two-loop MHV ratio $\cR^{(2)}_N$ known for any number of particles at LLA \cite{Bartels:2011ge} and NLLA \cite{Bargheer:2015djt,DelDuca:2018hrv}:
\begin{align}
\tilde{g}_{++++}^{(2;0,0,0)} &= \frac{\cR^{(2)}_8 - i \delta^{(2)}_8}{2 \pi i} = \tilde{g}_{\varpi;++++}^{(2;0,0,0)} \,, \\
\tilde{h}_{++++}^{(2;0,0,0)} &= \frac{(i \delta_8^{(1)})^2}{2(2 \pi i)^2} = \sum_{j=1}^{3}  {\tilde{h}}_{++++}^{j;(2;0,0,0)} + {\tilde{h}}_{\varpi;++++}^{(2;0,0,0)}\,.
\end{align}
We can lift these objects to three loops by convolution with the integration kernels $\cE_k$. Note however that due to the regularisation procedure at two loops given in \eqref{eq:def_regul}, the connection between the two- and the three loop perturbative coefficients is not as simple as described earlier. The regularized Fourier-Mellin integral at two loops consists of a combination of an integral transform with a modified contour and lower point terms that originate from the contour deformation. As only the former naturally obeys the convolution structure, we need to subtract these lower point contributions from the perturbative coefficients before inserting additional building blocks via convolutions.

Since performing this convolution preserves the contour of the original integral transform, the resulting expression with the inserted building block will have the same, deformed contour as the two-loop integral transform we started out from. This contour, however is not the same as the one in the Fourier-Mellin transform prescribed by $\cF_3$ at three loops. In order to obtain the desired building block we must add the corresponding three loop lower point contributions to restore the integration contour. Concretely, the relations between the two- and three loop corrected perturbative coefficients are given by
\beq\bsp
{\tilde{g}}_{\varpi;++++}^{(3;\delta_{k,1},\delta_{k,2},\delta_{k,3})} &= \cE(z_k)*\left({\tilde{g}}_{\varpi;++++}^{(2;0,0,0)}-P_{++++}^{(2;0,0,0)}\right) + P_{++++}^{(3;\delta_{k,1},\delta_{k,2},\delta_{k,3})},\\
{\tilde{h}}_{\varpi;++++}^{(3;\delta_{k,1},\delta_{k,2},\delta_{k,3})} &= \cE(z_k)*\left({\tilde{h}}_{\varpi;++++}^{(2;0,0,0)}-Q_{++++}^{(2;0,0,0)}\right) + Q_{++++}^{(3;\delta_{k,1},\delta_{k,2},\delta_{k,3})}.
\esp\eeq
where
\begin{align} \label{eq:def_P}
    P_{h_1h_2h_3 h_{4}}^{(\ell,i_1,i_2,i_{3})} &=  \tilde{g}^{(\ell,i_1,i_2+i_3)}_{h_1h_2h_4}(z_1,-z_2z_3)  +\delta_{i_1,0} \left(\tilde{g}^{(\ell;i_2,i_3)}_{h_2h_3h_4}(z_2,z_3) -  \tilde{g}^{(\ell;i_2+i_3)}_{h_2h_4}(-z_2z_3) \right),   \\
   Q_{h_1h_2h_3 h_{4}}^{(\ell,i_1,i_2,i_{3})} &=  \tilde{h}^{(\ell,i_1,i_2+i_3)}_{h_1h_2h_4}(z_1,-z_2z_3) + \delta_{i_1,0} \bigg[ \frac{1}{4} \cG_0(z_1)\tilde{g}^{(\ell-1;i_2,i_3)}_{h_2h_3h_4}(z_2,z_3)+\tilde{h}^{(\ell;i_2,i_3)}_{h_2h_3h_4}(z_2,z_3) \nonumber  \\
   &-   \left( \frac{1}{4} \cG_0(z_1) \tilde{g}^{(\ell-1;i_2+i_3)}_{h_2h_4}(-z_2z_3)+\tilde{h}^{(\ell;i_2+i_3)}_{h_2h_4}(-z_2z_3) \right) \bigg]. \label{eq:def_Q}
\end{align}
Note that as is the case for seven points \cite{DelDuca:2018hrv},
\begin{align}  \label{eq:Precur}
    P_{++++}^{(3;\delta_{k,1},\delta_{k,2},\delta_{k,3})} &=  \cE(z_k)*P_{++++}^{(2;0,0,0)} \,,\\
    Q_{++++}^{(3;\delta_{k,1},\delta_{k,2},\delta_{k,3})} &=  \cE(z_k)*\left(Q_{++++}^{(2;0,0,0)} - \frac{1}{32} \cG_0(z_k)^2\right)\,,
\end{align}
so that, in practice, we need only subtract lower point contributions for the real part.

From these MHV coefficients, we compute all helicity configurations by convoluting with the helicity flip kernel as explained in Section \ref{subsec:convo}.
In this case, at two loops no special precautions have to be made \cite{DelDuca:2018hrv} and we can directly perform convolutions with the helicity flip kernel on the full perturbative coefficients $\tilde{g}_{h_1h_2h_3h_4}^{(\ell;i_1,i_2,i_3)}$ and $\tilde{h}_{h_1h_2h_3h_4}^{(\ell;i_1,i_2,i_3)}$. At NLLA the helicity flip kernel contributes up to NLO
 and thus the computation consists of two types of contributions: a leading-order helicity flip kernel $\cH^{(0)}$ of the NLLA perturbative coefficient and a next-to-leading order helicity flip kernel $\cH^{(1)}$ of a LLA perturbative coefficient. The explicit form of the helicity flip kernel can be found in \eqref{eq:HKer}.
Flipping the helicity of the first radiated gluon
 for example we get
 \begin{equation}
 \begin{split}
 \tilde{g}_{-+++}^{(3;\delta_{k,1},\delta_{2,k},\delta_{k,3})}(\rho_1,\rho_2,\rho_3) =& \,\, \tilde{g}_{++++}^{(3;\delta_{k,1},\delta_{2,k},\delta_{k,3})}(\rho_1,\rho_2,\rho_3)*\cH^{(0)}(z_1)\\ &+ \tilde{g}_{++++}^{(2;\delta_{k,1},\delta_{2,k},\delta_{k,3})}(\rho_1,\rho_2,\rho_3)*\cH^{(1)}(z_1) \,,
 \end{split}
 \end{equation}
for $k \in \lbrace 1,2,3 \rbrace$. From this NMHV coefficient we can compute N$^2$MHV coefficients by performing further helicity flips, for example
\begin{equation}
\begin{split}
\tilde{g}_{--++}^{(3;\delta_{k,1},\delta_{2,k},\delta_{k,3})}(\rho_1,\rho_2,\rho_3) =& \,\, \tilde{g}_{-+++}^{(3;\delta_{k,1},\delta_{2,k},\delta_{k,3})}(\rho_1,\rho_2,\rho_3)*\cH^{(0)}(\bar{z}_1)*\cH^{(0)}(z_2)\\ &+ \tilde{g}_{-+++}^{(2;\delta_{k,1},\delta_{2,k},\delta_{k,3})}(\rho_1,\rho_2,\rho_3)*\cH^{(1)}(\bar{z}_1)*\cH^{(0)}(z_2)\\ &+ \tilde{g}_{-+++}^{(2;\delta_{k,1},\delta_{2,k},\delta_{k,3})}(\rho_1,\rho_2,\rho_3)*\cH^{(0)}(\bar{z}_1)*\cH^{(1)}(z_2) \,.
\end{split}
\end{equation}

Using the techniques outlined above we have computed the three loop amplitude at eight points in all helicity configurations up to next-to-leading logarithmic accuracy and we provide the corresponding perturbative coefficients in an ancillary file. The three loop MHV coefficient is also presented in appendix \ref{app:results}.

As expected, the amplitude \eqref{eq:MRK_conjecture} consist of linear combinations of the single-valued multiple polylogarithms introduced in section \ref{sec:polys}. In particular, all MHV perturbative coefficients are pure functions of uniform weight. Beyond MHV, the perturbative coefficients are linear combinations of polylogarithms with rational prefactors and we find that the leading singularity structure is equal to the one found at LLA in \cite{DelDuca:2016lad}. Up to three loops, the non-MHV ratio functions can therefore be expressed as

\begin{align}
\cR_{-+++} &= \fa^1 + \sum_{c=4}^{6} R_{23c} \fb^1_{c} \,, \\
\cR_{--++} &= \fa^2 + \sum_{b=2}^3 \sum_{b = 5}^6 R_{b4c} \fb^2_{bc} \,, \\
\cR_{+-++} &= \fa^3 + \bar{R}_{234} \fb^3_{4} + \sum_{c = 5}^6 R_{34c} \fb^3_{c} + \sum_{c = 5}^6 \bar{R}_{234}R_{34c} \fc^3_{c} \,, \\
\cR_{+--+} &= \fa^4 + \sum_{c=4}^5\bar{R}_{23c} \fb^4_{2c} + \sum_{b = 3}^4 R_{b56} \fb^4_{b6} + \sum_{b = 3}^4 \sum_{c=4}^5 \bar{R}_{23c} R_{b56} \fc^4_{bc} \,, \\
\cR_{-+-+} &= \fa^5 + R_{456} \fb^5_{006} + \sum_{c_1=4}^6 R_{23c_1} \fb^5_{c_100} + \sum_{c_2=5}^6 \bar{R}_{34c_2} \fb^5_{0c_20} \\
				   &+ \sum_{c_1=4}^6 R_{456} R_{23c_1} \fc_{c_106}^{5} + \sum_{c_2=5}^6 \bar{R}_{34c_2} R_{456} \fc_{0c_26}^{5}  \nonumber \\
				   &+ \sum_{c_1=4}^6  \sum_{c_2=5}^6 R_{23c_1} \bar{R}_{34c_2} \fc_{c_1c_20}^{5}+ \sum_{c_1=4}^6  \sum_{c_2=5}^6 R_{23c_1} \bar{R}_{34c_2} R_{456} \fd_{c_1c_26}^5 \,, \nonumber
\end{align}
where
\begin{equation}
R_{bac} = \frac{(\bx_b - \bx_a)(\bx_c - 1)}{(\bx_b - \bx_c)(\bx_a - 1)},
\end{equation}
$\bar{R}_{bac}$ are their complex conjugates and the $\fa, \fb, \fc, \fd$ are pure linear combinations of SVMPLs. Note that the cross ratios $R_{bac}$ are not independent, but they satisfy intricate non-linear relations as described in \cite{DelDuca:2016lad}.

We have verified that under soft limits, when expressed in terms of the Fourier-Mellin coordinates via \eqref{ztorho}, the computed perturbative coefficients evaluate to the correct combination of seven-point perturbative coefficients such that the equations \eqref{eq:soft1}-\eqref{eq:soft4} are fulfilled.

\section{Conclusion}\label{sec:conclu}

In \cite{DelDuca:2016lad} a mathematical framework was introduced to efficiently compute scattering amplitudes in the multi-Regge-limit of planar $\cN = 4$ super-Yang-Mills theory at leading logarithmic accuracy, which was later extended to the six- and seven-gluon amplitudes at next-to-leading logarithmic accuracy \cite{DelDuca:2018hrv} and will be generalised to any number of particles in \cite{FutureCite}. In this paper we applied the developed formalism to eight-gluon amplitudes, where the dispersion integral contains for the first time multiple central emission vertices. These objects which appear first for seven particles are the only building blocks exhibiting both a real and imaginary part and are the reason why the real part of the amplitude is not completely determined by LLA perturbative coefficients as is the case for six-gluon scattering \cite{Dixon:2012yy}.

Based on the work done in the seven particle case in \cite{DelDuca:2018hrv}, we worked out a well-defined Fourier-Mellin representation of the all order dispersion integral \eqref{eq:contourFixN} for eight external particles that allows for the expansion at weak coupling, assuming that the factorization into BFKL building blocks holds to any number of particles. This Fourier-Mellin representation lies at the heart of our formalism, as it allows us to relate perturbative coefficients at different loop orders through convolution integrals.

We have computed the perturbative coefficients needed to describe the eight-gluon amplitude at NLLA through three loops in any helicity configuration and provide our results in the ancillary files \texttt{gTilde.m} and \texttt{hTilde.m} in \textsc{Mathematica} format. Using the factorization theorem beyond LLA, which will be given in \cite{FutureCite}, the computed perturbative coefficients capture the NLLA three loop amplitudes in the MHV configuration for any multiplicity.

\section*{Acknowledgements}

We thank Vittorio Del Duca, Stefan Druc, James Drummond, Claude Duhr, Falko Dulat and Georgios Papathanasiou for collaboration in early stages of the project. We thank Vittorio Del Duca, James Drummond, Claude Duhr, Falko Dulat and Georgios Papathanasiou for their comments on the manuscript. We are grateful for the hospitality of the CERN TH department and the GGI Institute for Theoretical Physics during various stages of this work. This work is supported by the European Research Council (ERC) under the Horizon 2020 Research and Innovation Programme through the grant 637019 (MathAm).

\appendix

\section{BFKL Building Blocks}\label{app:BBB}
In this appendix we show the BFKL building blocks from \eqref{eq:MRK_conjecture} up to NLO explicitly. The impact factor and BFKL eigenvalue were given to all orders in \cite{Basso:2014pla} and up to NLO they are given by
\begin{align}
\chi^+(\nu,n)&=\frac{1}{\nu-\frac{i n}{2}}\left[1-\frac{a}{4}\left(E^2+\frac{3}{4}N^2-NV+\frac{\pi^2}{3}\right)+\cO(a^2)\right]\,,\\
\chi^-(\nu,n)&=\frac{1}{\nu+\frac{i n}{2}}\left[1-\frac{a}{4}\left(E^2+\frac{3}{4}N^2+NV+\frac{\pi^2}{3}\right)+\cO(a^2)\right]\,,\\
-\omega(\nu,n)  =&\, \,  a E - \frac{a^2}{4}\left(D^2E-2VDE+4\zeta_2 E+12\zeta_3\right) + \mathcal{O}(a^{3})\,.
\end{align}
The central emission block at NLO was first derived in \cite{DelDuca:2018hrv} and is given by
\beq\bsp
C^+(\nu_1,n_1,\nu_2,n_2) =&\frac{-\Gamma \bigl(1-i \nu_1 - \tfrac{n_1}{2}\bigr)\Gamma \bigl(i \nu_2 + \tfrac{n_2}{2}\bigr) \Gamma \bigl(i \nu_1 - i \nu_2 - \tfrac{n_1}{2} + \tfrac{n_2}{2}\bigr)}{\Gamma \bigl(1+i \nu_1 - \tfrac{n_1}{2}\bigr) \Gamma \bigl(- i \nu_2 + \tfrac{n_2}{2}\bigr) \Gamma \bigl(1- i \nu_1 + i \nu_2 - \tfrac{n_1}{2} + \tfrac{n_2}{2}\bigr)}\\
&\times\bigg[1+a\big(\frac{1}{2}\big[ DE_1-DE_2   + E_1 E_2  +\tfrac{1}{4} (N_1+ N_2)^2 + V_1 V_2\\
&+ (V_1-V_2)\bigl(M-E_1-E_2)+2\zeta_2+i\pi (V_2-V_1-E_1-E_2)\big]\\
&-\tfrac{1}{4}(E_1^2+E_2^2+N_1 V_1-N_2 V_2)-\tfrac{3}{16}(N_1^2+N_2^2)-\zeta_2\big)+\cO(a^2)\bigg].
\esp\eeq
Where the building blocks are themselves built out of combinations of the objects
\beq\bsp
 E(\nu,n)&=-\frac12\frac{|n|}{\nu^2+\frac{n^2}{4}}+\psi\left(1+i\nu+\frac{|n|}{2}\right) +\psi\left(1-i\nu+\frac{|n|}{2}\right)-2\psi(1)\,,
\\
V(\nu,n)&= \frac{i\nu}{\nu^2+\frac{n^2}{4}}\,,\qquad N(\nu,n) = \frac{n}{\nu^2+\frac{n^2}{4}}\,,
\qquad D_\nu=-i\partial/\partial \nu\,,\\
M(\nu_1,n_1,\nu_2,n_2)&=\psi(i(\nu_1-\nu_2)-\tfrac{n_1-n_2}{2}) + \psi(1-i(\nu_1-\nu_2)-\tfrac{n_1-n_2}{2})-2\psi(1)\,.
\esp\eeq

\section{Explicit Result}\label{app:results}

In this section we present explicitly the independent three-loop MHV perturbative coefficients at NLLA. Using target-projectile symmetry, these yield the full three-loop MHV remainder function. For compactness, we introduce the shorthand
\begin{equation}
\cG_{\vec{a}}^i \equiv \cG_{\vec{a}}(\rho_i).
\end{equation}
\begin{align}
\tilde{h}_{++++}^{(1,0,0)} \left( \rho_1, \rho_2, \rho_3 \right) = &-\frac{1}{16}\cG_{0,0,1}^1-\frac{1}{8}\cG_{0,1,0}^1+\frac{1}{8}\cG_{0,1,1}^1-\frac{1}{16}\cG_{1,0,0}^1+\frac{1}{8}\cG_{1,0,1}^1+\frac{1}{8}\cG_{1,1,0}^1\\
 \nonumber &-\frac{1}{8}\cG_{1,1,1}^1+\frac{1}{16}\cG_1^3\cG_{0,1}^1+\frac{1}{16}\cG_1^3\cG_{1,0}^1-\frac{1}{8}\cG_1^3\cG_{1,1}^1 \,.
\end{align}
\begin{align}
\tilde{h}_{++++}^{(0,1,0)} \left( \rho_1, \rho_2, \rho_3 \right) = &\frac{1}{16}\cG_{0,0,1}^2-\frac{1}{8}\cG_{0,1,1}^2+\frac{1}{16}\cG_{1,0,0}^2-\frac{1}{8}\cG_{1,0,1}^2-\frac{1}{8}\cG_{1,1,0}^2+\frac{1}{4}\cG_{1,1,1}^2-\frac{\zeta_3}{8}\\
 \nonumber &-\frac{1}{16}\cG_0^1\cG_{0,1}^2+\frac{1}{16}\cG_1^1\cG_{0,1}^2+\frac{1}{16}\cG_1^3\cG_{0,1}^2-\frac{1}{16}\cG_0^1\cG_{1,0}^2+\frac{1}{16}\cG_1^1\cG_{1,0}^2\\
 \nonumber &+\frac{1}{16}\cG_1^3\cG_{1,0}^2+\frac{1}{8}\cG_0^1\cG_{1,1}^2-\frac{1}{8}\cG_1^1\cG_{1,1}^2-\frac{1}{8}\cG_1^3\cG_{1,1}^2 \,.
\end{align}

\begin{align}
\tilde{g}_{++++}^{(1,0,0)} \left( \rho_1, \rho_2, \rho_3 \right) = &\frac{3}{16}\cG_{0,0,0,1}^1-\frac{1}{16}\cG_{0,0,1,0}^1-\frac{3}{8}\cG_{0,0,1,1}^1+\frac{1}{16}\cG_{0,0,1,\rho_2}^1-\frac{1}{16}\cG_{0,0,\rho_2,1}^1\\
 \nonumber &-\frac{1}{16}\cG_{0,1,0,0}^1-\frac{1}{4}\cG_{0,1,0,1}^1+\frac{3}{4}\cG_{0,1,1,1}^1+\frac{1}{8}\cG_{0,1,\rho_2,\rho_2}^1-\frac{1}{8}\cG_{0,1,\rho_3,\rho_2}^1\\
 \nonumber &+\frac{1}{8}\cG_{0,1,\rho_3,\rho_3}^1+\frac{1}{16}\cG_{0,\rho_2,0,1}^1+\frac{1}{4}\cG_{0,\rho_2,1,1}^1-\frac{1}{8}\cG_{0,\rho_2,1,\rho_2}^1-\frac{1}{8}\cG_{0,\rho_2,\rho_3,1}^1\\
 \nonumber &+\frac{1}{8}\cG_{0,\rho_3,\rho_3,1}^1+\frac{3}{16}\cG_{1,0,0,0}^1-\frac{1}{2}\cG_{1,0,0,1}^1-\frac{1}{16}\cG_{1,0,0,\rho_2}^1-\frac{1}{4}\cG_{1,0,1,0}^1\\
 \nonumber &+\frac{3}{4}\cG_{1,0,1,1}^1+\frac{1}{16}\cG_{1,0,1,\rho_2}^1+\frac{1}{16}\cG_{1,0,\rho_2,0}^1+\frac{1}{16}\cG_{1,0,\rho_2,1}^1-\frac{3}{8}\cG_{1,1,0,0}^1\\
 \nonumber &+\frac{3}{4}\cG_{1,1,0,1}^1+\frac{1}{8}\cG_{1,1,0,\rho_2}^1+\frac{3}{4}\cG_{1,1,1,0}^1-\frac{3}{2}\cG_{1,1,1,1}^1-\frac{3}{16}\cG_{1,1,1,\rho_2}^1\\
 \nonumber &+\frac{1}{4}\cG_{1,1,\rho_2,0}^1-\frac{5}{16}\cG_{1,1,\rho_2,1}^1-\frac{1}{8}\cG_{1,1,\rho_2,\rho_2}^1+\frac{1}{8}\cG_{1,1,\rho_3,\rho_2}^1-\frac{1}{8}\cG_{1,1,\rho_3,\rho_3}^1\\
 \nonumber &-\frac{1}{16}\cG_{1,\rho_2,0,0}^1+\frac{1}{16}\cG_{1,\rho_2,0,1}^1+\frac{1}{16}\cG_{1,\rho_2,0,\rho_2}^1-\frac{5}{16}\cG_{1,\rho_2,1,1}^1+\frac{1}{16}\cG_{1,\rho_2,1,\rho_2}^1\\
 \nonumber &+\frac{1}{8}\cG_{1,\rho_2,\rho_3,1}^1-\frac{1}{8}\cG_{1,\rho_3,\rho_2,0}^1+\frac{1}{8}\cG_{1,\rho_3,\rho_2,1}^1+\frac{1}{8}\cG_{1,\rho_3,\rho_3,0}^1-\frac{1}{4}\cG_{1,\rho_3,\rho_3,1}^1\\
 \nonumber &-\frac{1}{16}\cG_{\rho_2,0,0,1}^1+\frac{1}{8}\cG_{\rho_2,0,1,1}^1-\frac{1}{16}\cG_{\rho_2,0,1,\rho_2}^1+\frac{1}{16}\cG_{\rho_2,0,\rho_2,1}^1+\frac{1}{16}\cG_{\rho_2,1,0,0}^1\\
 \nonumber &+\frac{1}{16}\cG_{\rho_2,1,0,1}^1-\frac{1}{16}\cG_{\rho_2,1,0,\rho_2}^1-\frac{3}{16}\cG_{\rho_2,1,1,1}^1+\frac{1}{8}\cG_{\rho_2,1,1,\rho_2}^1-\frac{1}{8}\cG_{\rho_2,1,\rho_2,0}^1\\
 \nonumber &+\frac{1}{16}\cG_{\rho_2,1,\rho_2,1}^1+\frac{1}{8}\cG_{\rho_2,\rho_2,1,0}^1-\frac{1}{8}\cG_{\rho_2,\rho_2,1,1}^1-\frac{1}{8}\cG_{\rho_2,\rho_3,1,0}^1+\frac{1}{8}\cG_{\rho_2,\rho_3,1,1}^1\\
 \nonumber &+\frac{1}{8}\cG_{\rho_3,\rho_3,1,0}^1-\frac{1}{8}\cG_{\rho_3,\rho_3,1,1}^1+\frac{1}{2}\cG_{0,1}^1\zeta_2+\frac{1}{2}\cG_{1,0}^1\zeta_2-\cG_{1,1}^1\zeta_2-\cG_1^1\frac{3}{4}\zeta_3\\
 \nonumber &+\frac{1}{8}\cG_{0,0}^2\cG_{0,1}^1+\frac{1}{8}\cG_{0,0}^3\cG_{0,1}^1+\frac{1}{16}\cG_{0,1}^1\cG_{0,1}^2-\frac{1}{16}\cG_{0,1}^1\cG_{0,1}^3-\frac{3}{16}\cG_{0,1}^2\cG_{0,\rho_2}^1\\
 \nonumber &-\frac{1}{16}\cG_{0,1}^3\cG_{0,\rho_3}^1-\frac{1}{16}\cG_{0,1}^1\cG_{0,\rho_3}^2+\frac{1}{8}\cG_{0,0}^2\cG_{1,0}^1+\frac{1}{8}\cG_{0,0}^3\cG_{1,0}^1-\frac{1}{16}\cG_{0,1}^2\cG_{1,0}^1\\
 \nonumber &-\frac{1}{16}\cG_{0,1}^3\cG_{1,0}^1-\frac{1}{16}\cG_{1,0}^1\cG_{0,\rho_3}^2-\frac{3}{16}\cG_{0,1}^1\cG_{1,0}^2+\frac{1}{16}\cG_{1,0}^2\cG_{0,\rho_2}^1-\frac{3}{16}\cG_{1,0}^1\cG_{1,0}^2\\
 \nonumber &-\frac{3}{16}\cG_{0,1}^1\cG_{1,0}^3+\frac{1}{16}\cG_{1,0}^3\cG_{0,\rho_3}^1-\frac{3}{16}\cG_{1,0}^1\cG_{1,0}^3-\frac{1}{4}\cG_{0,0}^2\cG_{1,1}^1-\frac{1}{8}\cG_{0,0}^3\cG_{1,1}^1\\
 \nonumber &+\frac{3}{16}\cG_{0,1}^2\cG_{1,1}^1+\frac{1}{16}\cG_{0,1}^3\cG_{1,1}^1+\frac{1}{16}\cG_{1,1}^1\cG_{0,\rho_3}^2+\frac{3}{16}\cG_{1,0}^2\cG_{1,1}^1+\frac{3}{16}\cG_{1,0}^3\cG_{1,1}^1\\
 \nonumber &+\frac{1}{4}\cG_{1,1}^2\cG_{0,\rho_2}^1+\frac{1}{8}\cG_{1,0}^1\cG_{1,1}^2-\frac{1}{8}\cG_{1,1}^1\cG_{1,1}^2+\frac{1}{8}\cG_{0,1}^1\cG_{1,1}^3+\frac{1}{8}\cG_{1,1}^3\cG_{0,\rho_3}^1\\
 \nonumber &+\frac{1}{8}\cG_{1,0}^1\cG_{1,1}^3-\frac{1}{8}\cG_{1,1}^1\cG_{1,1}^3-\frac{3}{16}\cG_{0,0}^2\cG_{1,\rho_2}^1+\frac{3}{16}\cG_{0,1}^2\cG_{1,\rho_2}^1+\frac{1}{16}\cG_{0,\rho_3}^2\cG_{1,\rho_2}^1\\
 \nonumber &+\frac{1}{8}\cG_{1,0}^2\cG_{1,\rho_2}^1-\frac{1}{4}\cG_{1,1}^2\cG_{1,\rho_2}^1+\frac{1}{8}\cG_{0,0}^2\cG_{1,\rho_3}^1-\frac{1}{4}\cG_{0,0}^3\cG_{1,\rho_3}^1-\frac{1}{8}\cG_{0,1}^2\cG_{1,\rho_3}^1\\
 \nonumber &+\frac{1}{4}\cG_{0,1}^3\cG_{1,\rho_3}^1+\frac{1}{8}\cG_{1,0}^3\cG_{1,\rho_3}^1-\frac{1}{4}\cG_{1,1}^3\cG_{1,\rho_3}^1+\frac{1}{16}\cG_{0,1}^1\cG_{1,\rho_3}^2-\frac{1}{16}\cG_{0,\rho_2}^1\cG_{1,\rho_3}^2\\
 \nonumber &+\frac{1}{16}\cG_{1,0}^1\cG_{1,\rho_3}^2-\frac{1}{16}\cG_{1,1}^1\cG_{1,\rho_3}^2-\frac{1}{16}\cG_{0,1}^2\cG_{\rho_2,0}^1+\frac{1}{16}\cG_{1,0}^2\cG_{\rho_2,0}^1+\frac{1}{8}\cG_{1,1}^2\cG_{\rho_2,0}^1
 \end{align}
 \begin{align}
 \nonumber &-\frac{1}{16}\cG_{1,\rho_3}^2\cG_{\rho_2,0}^1+\frac{1}{16}\cG_{0,0}^2\cG_{\rho_2,1}^1-\frac{1}{8}\cG_{0,1}^2\cG_{\rho_2,1}^1+\frac{1}{8}\cG_{0,1}^3\cG_{\rho_2,1}^1-\frac{1}{16}\cG_{1,0}^2\cG_{\rho_2,1}^1 \\
 \nonumber &-\frac{1}{8}\cG_{1,1}^3\cG_{\rho_2,1}^1+\frac{1}{16}\cG_{1,\rho_3}^2\cG_{\rho_2,1}^1+\frac{1}{8}\cG_{0,1}^2\cG_{\rho_2,\rho_2}^1-\frac{1}{8}\cG_{1,1}^2\cG_{\rho_2,\rho_2}^1-\frac{1}{8}\cG_{0,1}^3\cG_{\rho_2,\rho_3}^1\\
 \nonumber &+\frac{1}{8}\cG_{1,1}^3\cG_{\rho_2,\rho_3}^1-\frac{1}{16}\cG_{0,1}^3\cG_{\rho_3,0}^1+\frac{1}{16}\cG_{1,0}^3\cG_{\rho_3,0}^1+\frac{1}{8}\cG_{1,1}^3\cG_{\rho_3,0}^1+\frac{1}{16}\cG_{0,1}^1\cG_{\rho_3,0}^2\\
 \nonumber &+\frac{1}{16}\cG_{1,0}^1\cG_{\rho_3,0}^2-\frac{1}{16}\cG_{1,1}^1\cG_{\rho_3,0}^2+\frac{1}{16}\cG_{1,\rho_2}^1\cG_{\rho_3,0}^2-\frac{1}{8}\cG_{1,\rho_3}^1\cG_{\rho_3,0}^2-\frac{1}{16}\cG_{0,1}^3\cG_{\rho_3,1}^1\\
 \nonumber &-\frac{1}{16}\cG_{1,0}^3\cG_{\rho_3,1}^1-\frac{1}{16}\cG_{0,1}^1\cG_{\rho_3,1}^2+\frac{1}{16}\cG_{0,\rho_2}^1\cG_{\rho_3,1}^2-\frac{1}{16}\cG_{1,0}^1\cG_{\rho_3,1}^2+\frac{1}{16}\cG_{1,1}^1\cG_{\rho_3,1}^2\\
 \nonumber &-\frac{1}{8}\cG_{1,\rho_2}^1\cG_{\rho_3,1}^2+\frac{1}{8}\cG_{1,\rho_3}^1\cG_{\rho_3,1}^2+\frac{1}{16}\cG_{\rho_2,0}^1\cG_{\rho_3,1}^2-\frac{1}{16}\cG_{\rho_2,1}^1\cG_{\rho_3,1}^2+\frac{1}{8}\cG_{0,1}^3\cG_{\rho_3,\rho_3}^1\\
 \nonumber &-\frac{1}{8}\cG_{1,1}^3\cG_{\rho_3,\rho_3}^1-\frac{3}{16}\cG_1^1\cG_{0,0,0}^2-\frac{3}{16}\cG_1^1\cG_{0,0,0}^3+\frac{1}{16}\cG_0^2\cG_{0,0,1}^1-\frac{1}{16}\cG_1^2\cG_{0,0,1}^1\\
 \nonumber &+\frac{1}{8}\cG_{0,0,1}^2\cG_{\rho_2}^1+\frac{1}{8}\cG_{0,0,1}^3\cG_{\rho_3}^1+\frac{1}{8}\cG_1^2\cG_{0,0,\rho_2}^1+\frac{1}{16}\cG_1^1\cG_{0,0,\rho_3}^2+\frac{1}{4}\cG_1^1\cG_{0,1,0}^2\\
 \nonumber &-\frac{1}{16}\cG_{0,1,0}^2\cG_{\rho_2}^1+\frac{1}{4}\cG_1^1\cG_{0,1,0}^3-\frac{1}{16}\cG_{0,1,0}^3\cG_{\rho_3}^1+\frac{1}{16}\cG_1^1\cG_{0,1,1}^2-\frac{3}{16}\cG_{0,1,1}^2\cG_{\rho_2}^1\\
 \nonumber &+\frac{1}{16}\cG_1^1\cG_{0,1,1}^3-\frac{3}{16}\cG_{0,1,1}^3\cG_{\rho_3}^1+\frac{1}{8}\cG_0^2\cG_{0,1,\rho_2}^1-\frac{1}{16}\cG_0^3\cG_{0,1,\rho_2}^1-\frac{1}{4}\cG_1^2\cG_{0,1,\rho_2}^1\\
 \nonumber &+\frac{1}{16}\cG_1^3\cG_{0,1,\rho_2}^1-\frac{1}{16}\cG_{\rho_3}^2\cG_{0,1,\rho_2}^1-\frac{1}{8}\cG_0^2\cG_{0,1,\rho_3}^1+\frac{3}{16}\cG_0^3\cG_{0,1,\rho_3}^1+\frac{1}{16}\cG_1^2\cG_{0,1,\rho_3}^1\\
 \nonumber &-\frac{1}{8}\cG_1^3\cG_{0,1,\rho_3}^1+\frac{1}{16}\cG_{\rho_3}^2\cG_{0,1,\rho_3}^1-\frac{1}{8}\cG_1^1\cG_{0,1,\rho_3}^2+\frac{1}{16}\cG_{\rho_2}^1\cG_{0,1,\rho_3}^2+\frac{1}{8}\cG_1^2\cG_{0,\rho_2,0}^1\\
 \nonumber &-\frac{1}{8}\cG_0^2\cG_{0,\rho_2,1}^1+\frac{1}{16}\cG_0^3\cG_{0,\rho_2,1}^1-\frac{1}{4}\cG_1^2\cG_{0,\rho_2,1}^1-\frac{1}{16}\cG_1^3\cG_{0,\rho_2,1}^1+\frac{1}{16}\cG_{\rho_3}^2\cG_{0,\rho_2,1}^1\\
 \nonumber &+\frac{1}{8}\cG_1^3\cG_{0,\rho_2,\rho_3}^1+\frac{1}{16}\cG_1^1\cG_{0,\rho_3,0}^2-\frac{1}{16}\cG_0^3\cG_{0,\rho_3,1}^1+\frac{1}{16}\cG_1^2\cG_{0,\rho_3,1}^1-\frac{1}{16}\cG_{\rho_3}^2\cG_{0,\rho_3,1}^1\\
 \nonumber &-\frac{1}{16}\cG_{\rho_2}^1\cG_{0,\rho_3,1}^2-\frac{1}{8}\cG_1^3\cG_{0,\rho_3,\rho_3}^1-\frac{1}{16}\cG_0^2\cG_{1,0,0}^1+\frac{1}{16}\cG_1^2\cG_{1,0,0}^1+\frac{1}{8}\cG_1^1\cG_{1,0,0}^2\\
 \nonumber &+\frac{1}{8}\cG_1^1\cG_{1,0,0}^3+\frac{1}{8}\cG_0^2\cG_{1,0,1}^1-\frac{1}{8}\cG_1^2\cG_{1,0,1}^1-\frac{1}{16}\cG_1^1\cG_{1,0,1}^2-\frac{1}{16}\cG_{1,0,1}^2\cG_{\rho_2}^1\\
 \nonumber &-\frac{1}{16}\cG_1^1\cG_{1,0,1}^3-\frac{1}{16}\cG_{1,0,1}^3\cG_{\rho_3}^1-\frac{1}{16}\cG_0^2\cG_{1,0,\rho_2}^1-\frac{1}{8}\cG_1^2\cG_{1,0,\rho_2}^1+\frac{1}{4}\cG_0^2\cG_{1,1,0}^1\\
 \nonumber &-\frac{1}{4}\cG_1^2\cG_{1,1,0}^1-\frac{1}{16}\cG_1^1\cG_{1,1,0}^2+\frac{1}{16}\cG_{1,1,0}^2\cG_{\rho_2}^1-\frac{1}{16}\cG_1^1\cG_{1,1,0}^3+\frac{1}{16}\cG_{1,1,0}^3\cG_{\rho_3}^1\\
 \nonumber &-\frac{3}{16}\cG_0^2\cG_{1,1,1}^1+\frac{3}{16}\cG_1^2\cG_{1,1,1}^1-\frac{1}{8}\cG_1^1\cG_{1,1,1}^2+\frac{1}{8}\cG_{1,1,1}^2\cG_{\rho_2}^1-\frac{1}{8}\cG_1^1\cG_{1,1,1}^3\\
 \nonumber &+\frac{1}{8}\cG_{1,1,1}^3\cG_{\rho_3}^1-\frac{1}{4}\cG_0^2\cG_{1,1,\rho_2}^1+\frac{1}{16}\cG_0^3\cG_{1,1,\rho_2}^1+\frac{7}{16}\cG_1^2\cG_{1,1,\rho_2}^1-\frac{1}{16}\cG_1^3\cG_{1,1,\rho_2}^1\\
 \nonumber &+\frac{1}{16}\cG_{\rho_3}^2\cG_{1,1,\rho_2}^1+\frac{1}{8}\cG_0^2\cG_{1,1,\rho_3}^1-\frac{3}{16}\cG_0^3\cG_{1,1,\rho_3}^1-\frac{1}{16}\cG_1^2\cG_{1,1,\rho_3}^1+\frac{1}{8}\cG_1^3\cG_{1,1,\rho_3}^1\\
 \nonumber &-\frac{1}{16}\cG_{\rho_3}^2\cG_{1,1,\rho_3}^1+\frac{1}{16}\cG_1^1\cG_{1,1,\rho_3}^2-\frac{1}{16}\cG_{\rho_2}^1\cG_{1,1,\rho_3}^2+\frac{1}{8}\cG_0^2\cG_{1,\rho_2,0}^1-\frac{1}{16}\cG_0^3\cG_{1,\rho_2,0}^1\\
 \nonumber &-\frac{1}{8}\cG_1^2\cG_{1,\rho_2,0}^1+\frac{1}{16}\cG_1^3\cG_{1,\rho_2,0}^1-\frac{1}{16}\cG_{\rho_3}^2\cG_{1,\rho_2,0}^1+\frac{1}{4}\cG_1^2\cG_{1,\rho_2,1}^1-\frac{1}{8}\cG_1^3\cG_{1,\rho_2,\rho_3}^1\\
 \nonumber &-\frac{1}{8}\cG_0^2\cG_{1,\rho_3,0}^1+\frac{3}{16}\cG_0^3\cG_{1,\rho_3,0}^1+\frac{1}{16}\cG_1^2\cG_{1,\rho_3,0}^1-\frac{1}{8}\cG_1^3\cG_{1,\rho_3,0}^1+\frac{1}{16}\cG_{\rho_3}^2\cG_{1,\rho_3,0}^1\\
 \nonumber &-\frac{1}{16}\cG_1^1\cG_{1,\rho_3,1}^2+\frac{1}{16}\cG_{\rho_2}^1\cG_{1,\rho_3,1}^2+\frac{1}{8}\cG_0^2\cG_{1,\rho_3,\rho_2}^1-\frac{1}{8}\cG_1^2\cG_{1,\rho_3,\rho_2}^1-\frac{1}{8}\cG_0^3\cG_{1,\rho_3,\rho_3}^1
 \end{align}
 \begin{align}
 \nonumber &+\frac{1}{4}\cG_1^3\cG_{1,\rho_3,\rho_3}^1-\frac{1}{16}\cG_0^2\cG_{\rho_2,0,1}^1-\frac{1}{8}\cG_0^2\cG_{\rho_2,1,0}^1+\frac{1}{16}\cG_0^3\cG_{\rho_2,1,0}^1-\frac{1}{16}\cG_1^3\cG_{\rho_2,1,0}^1\\
 \nonumber &+\frac{1}{16}\cG_{\rho_3}^2\cG_{\rho_2,1,0}^1+\frac{1}{8}\cG_0^2\cG_{\rho_2,1,1}^1-\frac{1}{16}\cG_0^3\cG_{\rho_2,1,1}^1+\frac{1}{16}\cG_1^2\cG_{\rho_2,1,1}^1+\frac{1}{16}\cG_1^3\cG_{\rho_2,1,1}^1\\
 \nonumber &-\frac{1}{16}\cG_{\rho_3}^2\cG_{\rho_2,1,1}^1+\frac{1}{16}\cG_0^2\cG_{\rho_2,1,\rho_2}^1-\frac{1}{16}\cG_1^2\cG_{\rho_2,1,\rho_2}^1-\frac{1}{8}\cG_1^2\cG_{\rho_2,\rho_2,0}^1+\frac{1}{8}\cG_1^2\cG_{\rho_2,\rho_2,1}^1\\
 \nonumber &+\frac{1}{8}\cG_1^3\cG_{\rho_2,\rho_3,0}^1-\frac{1}{8}\cG_1^3\cG_{\rho_2,\rho_3,1}^1-\frac{1}{16}\cG_1^1\cG_{\rho_3,0,0}^2+\frac{1}{8}\cG_1^1\cG_{\rho_3,0,1}^2-\frac{1}{16}\cG_{\rho_2}^1\cG_{\rho_3,0,1}^2\\
 \nonumber &-\frac{1}{16}\cG_0^3\cG_{\rho_3,1,0}^1+\frac{1}{16}\cG_1^2\cG_{\rho_3,1,0}^1-\frac{1}{16}\cG_{\rho_3}^2\cG_{\rho_3,1,0}^1+\frac{1}{16}\cG_0^3\cG_{\rho_3,1,1}^1-\frac{1}{16}\cG_1^2\cG_{\rho_3,1,1}^1\\
 \nonumber &+\frac{1}{16}\cG_{\rho_3}^2\cG_{\rho_3,1,1}^1-\frac{1}{16}\cG_1^1\cG_{\rho_3,1,1}^2+\frac{1}{16}\cG_{\rho_2}^1\cG_{\rho_3,1,1}^2-\frac{1}{8}\cG_1^3\cG_{\rho_3,\rho_3,0}^1+\frac{1}{8}\cG_1^3\cG_{\rho_3,\rho_3,1}^1\\
 \nonumber &+\frac{1}{16}\cG_0^3\cG_1^1\cG_{0,0}^2-\frac{1}{16}\cG_1^1\cG_1^3\cG_{0,0}^2-\frac{1}{16}\cG_1^1\cG_1^2\cG_{0,0}^3+\frac{1}{16}\cG_1^1\cG_{0,0}^3\cG_{\rho_3}^2-\frac{1}{16}\cG_0^2\cG_0^3\cG_{0,1}^1\\
 \nonumber &+\frac{1}{8}\cG_0^3\cG_1^2\cG_{0,1}^1+\frac{1}{16}\cG_0^2\cG_1^3\cG_{0,1}^1-\frac{1}{8}\cG_1^2\cG_1^3\cG_{0,1}^1-\frac{1}{16}\cG_0^3\cG_{0,1}^1\cG_{\rho_3}^2+\frac{1}{16}\cG_1^3\cG_{0,1}^1\cG_{\rho_3}^2\\
 \nonumber &-\frac{1}{8}\cG_0^3\cG_1^1\cG_{0,1}^2+\frac{1}{8}\cG_1^1\cG_1^3\cG_{0,1}^2+\frac{1}{16}\cG_0^3\cG_{0,1}^2\cG_{\rho_2}^1-\frac{1}{16}\cG_1^3\cG_{0,1}^2\cG_{\rho_2}^1+\frac{1}{8}\cG_1^1\cG_1^2\cG_{0,1}^3\\
 \nonumber &-\frac{1}{8}\cG_1^2\cG_{0,1}^3\cG_{\rho_2}^1+\frac{1}{16}\cG_1^2\cG_{0,1}^3\cG_{\rho_3}^1-\frac{1}{8}\cG_1^1\cG_{0,1}^3\cG_{\rho_3}^2+\frac{1}{8}\cG_{0,1}^3\cG_{\rho_2}^1\cG_{\rho_3}^2-\frac{1}{16}\cG_{0,1}^3\cG_{\rho_3}^1\cG_{\rho_3}^2\\
 \nonumber &-\frac{1}{16}\cG_0^3\cG_1^2\cG_{0,\rho_2}^1+\frac{1}{16}\cG_1^2\cG_1^3\cG_{0,\rho_2}^1-\frac{1}{8}\cG_1^3\cG_{\rho_3}^2\cG_{0,\rho_2}^1-\frac{1}{16}\cG_1^2\cG_1^3\cG_{0,\rho_3}^1+\frac{1}{16}\cG_1^3\cG_{\rho_3}^2\cG_{0,\rho_3}^1\\
 \nonumber &-\frac{1}{8}\cG_1^1\cG_1^3\cG_{0,\rho_3}^2+\frac{1}{8}\cG_1^3\cG_{\rho_2}^1\cG_{0,\rho_3}^2-\frac{1}{16}\cG_0^2\cG_0^3\cG_{1,0}^1+\frac{1}{8}\cG_0^3\cG_1^2\cG_{1,0}^1+\frac{1}{16}\cG_0^2\cG_1^3\cG_{1,0}^1\\
 \nonumber &-\frac{1}{8}\cG_1^2\cG_1^3\cG_{1,0}^1-\frac{1}{16}\cG_0^3\cG_{1,0}^1\cG_{\rho_3}^2+\frac{1}{16}\cG_1^3\cG_{1,0}^1\cG_{\rho_3}^2+\frac{1}{16}\cG_0^2\cG_0^3\cG_{1,1}^1-\frac{1}{8}\cG_0^3\cG_1^2\cG_{1,1}^1\\
 \nonumber &-\frac{1}{16}\cG_0^2\cG_1^3\cG_{1,1}^1+\frac{1}{8}\cG_1^2\cG_1^3\cG_{1,1}^1+\frac{1}{16}\cG_0^3\cG_{1,1}^1\cG_{\rho_3}^2-\frac{1}{16}\cG_1^3\cG_{1,1}^1\cG_{\rho_3}^2+\frac{1}{16}\cG_0^3\cG_1^1\cG_{1,1}^2\\
 \nonumber &-\frac{1}{16}\cG_1^1\cG_1^3\cG_{1,1}^2-\frac{1}{16}\cG_0^3\cG_{1,1}^2\cG_{\rho_2}^1+\frac{1}{16}\cG_1^3\cG_{1,1}^2\cG_{\rho_2}^1-\frac{1}{16}\cG_1^1\cG_1^2\cG_{1,1}^3+\frac{1}{8}\cG_1^2\cG_{1,1}^3\cG_{\rho_2}^1\\
 \nonumber &-\frac{1}{16}\cG_1^2\cG_{1,1}^3\cG_{\rho_3}^1+\frac{1}{16}\cG_1^1\cG_{1,1}^3\cG_{\rho_3}^2-\frac{1}{8}\cG_{1,1}^3\cG_{\rho_2}^1\cG_{\rho_3}^2+\frac{1}{16}\cG_{1,1}^3\cG_{\rho_3}^1\cG_{\rho_3}^2+\frac{1}{16}\cG_0^2\cG_0^3\cG_{1,\rho_2}^1\\
 \nonumber &-\frac{1}{16}\cG_0^2\cG_1^3\cG_{1,\rho_2}^1+\frac{1}{8}\cG_1^3\cG_{\rho_3}^2\cG_{1,\rho_2}^1-\frac{1}{16}\cG_0^3\cG_1^2\cG_{1,\rho_3}^1+\frac{1}{8}\cG_1^2\cG_1^3\cG_{1,\rho_3}^1+\frac{1}{16}\cG_0^3\cG_{\rho_3}^2\cG_{1,\rho_3}^1\\
 \nonumber &-\frac{1}{8}\cG_1^3\cG_{\rho_3}^2\cG_{1,\rho_3}^1+\frac{1}{8}\cG_1^1\cG_1^3\cG_{1,\rho_3}^2-\frac{1}{8}\cG_1^3\cG_{\rho_2}^1\cG_{1,\rho_3}^2-\frac{1}{16}\cG_0^3\cG_1^2\cG_{\rho_2,0}^1+\frac{1}{16}\cG_1^2\cG_1^3\cG_{\rho_2,0}^1\\
 \nonumber &-\frac{1}{8}\cG_1^3\cG_{\rho_3}^2\cG_{\rho_2,0}^1+\frac{1}{16}\cG_0^3\cG_1^2\cG_{\rho_2,1}^1-\frac{1}{16}\cG_1^2\cG_1^3\cG_{\rho_2,1}^1+\frac{1}{8}\cG_1^3\cG_{\rho_3}^2\cG_{\rho_2,1}^1-\frac{1}{16}\cG_1^2\cG_1^3\cG_{\rho_3,0}^1\\
 \nonumber &+\frac{1}{16}\cG_1^3\cG_{\rho_3}^2\cG_{\rho_3,0}^1+\frac{1}{16}\cG_1^2\cG_1^3\cG_{\rho_3,1}^1-\frac{1}{16}\cG_1^3\cG_{\rho_3}^2\cG_{\rho_3,1}^1\,.
\end{align}
\begin{align}
\tilde{g}_{++++}^{(0,1,0)} \left( \rho_1, \rho_2, \rho_3 \right) = &\frac{3}{16}\cG_{0,0,0,1}^2-\frac{1}{16}\cG_{0,0,1,0}^2-\frac{3}{8}\cG_{0,0,1,1}^2+\frac{1}{8}\cG_{0,0,1,\rho_2}^1+\frac{1}{16}\cG_{0,0,1,\rho_3}^2\\
 \nonumber &+\frac{1}{8}\cG_{0,0,\rho_2,1}^1-\frac{1}{16}\cG_{0,0,\rho_3,1}^2-\frac{1}{16}\cG_{0,1,0,0}^2-\frac{1}{4}\cG_{0,1,0,1}^2-\frac{1}{16}\cG_{0,1,0,\rho_2}^1\\
 \nonumber &+\frac{3}{4}\cG_{0,1,1,1}^2-\frac{3}{16}\cG_{0,1,1,\rho_2}^1-\frac{1}{8}\cG_{0,1,\rho_2,0}^1-\frac{1}{4}\cG_{0,1,\rho_2,1}^1+\frac{1}{8}\cG_{0,1,\rho_2,\rho_2}^1\\
 \nonumber &+\frac{1}{8}\cG_{0,1,\rho_3,\rho_3}^2-\frac{1}{16}\cG_{0,\rho_2,0,1}^1-\frac{1}{8}\cG_{0,\rho_2,1,0}^1-\frac{1}{16}\cG_{0,\rho_2,1,1}^1+\frac{1}{16}\cG_{0,\rho_3,0,1}^2\\
 \nonumber &+\frac{1}{4}\cG_{0,\rho_3,1,1}^2-\frac{1}{8}\cG_{0,\rho_3,1,\rho_3}^2+\frac{3}{16}\cG_{1,0,0,0}^2-\frac{1}{2}\cG_{1,0,0,1}^2-\frac{1}{16}\cG_{1,0,0,\rho_3}^2\\
 \nonumber &-\frac{1}{4}\cG_{1,0,1,0}^2+\frac{3}{4}\cG_{1,0,1,1}^2-\frac{1}{16}\cG_{1,0,1,\rho_2}^1+\frac{1}{16}\cG_{1,0,1,\rho_3}^2-\frac{1}{16}\cG_{1,0,\rho_2,0}^1\\
 \nonumber &+\frac{1}{16}\cG_{1,0,\rho_3,0}^2+\frac{1}{16}\cG_{1,0,\rho_3,1}^2-\frac{3}{8}\cG_{1,1,0,0}^2+\frac{3}{4}\cG_{1,1,0,1}^2+\frac{1}{16}\cG_{1,1,0,\rho_2}^1\\
 \nonumber &+\frac{1}{8}\cG_{1,1,0,\rho_3}^2+\frac{3}{4}\cG_{1,1,1,0}^2-\frac{3}{2}\cG_{1,1,1,1}^2+\frac{1}{8}\cG_{1,1,1,\rho_2}^1-\frac{3}{16}\cG_{1,1,1,\rho_3}^2\\
 \nonumber &-\frac{1}{16}\cG_{1,1,\rho_2,0}^1+\frac{3}{8}\cG_{1,1,\rho_2,1}^1-\frac{1}{8}\cG_{1,1,\rho_2,\rho_2}^1+\frac{1}{4}\cG_{1,1,\rho_3,0}^2-\frac{5}{16}\cG_{1,1,\rho_3,1}^2\\
 \nonumber &-\frac{1}{8}\cG_{1,1,\rho_3,\rho_3}^2+\frac{1}{8}\cG_{1,\rho_2,0,0}^1-\frac{1}{16}\cG_{1,\rho_2,0,\rho_2}^1-\frac{1}{4}\cG_{1,\rho_2,1,0}^1+\frac{3}{8}\cG_{1,\rho_2,1,1}^1\\
 \nonumber &+\frac{1}{16}\cG_{1,\rho_2,1,\rho_2}^1-\frac{1}{16}\cG_{1,\rho_3,0,0}^2+\frac{1}{16}\cG_{1,\rho_3,0,1}^2+\frac{1}{16}\cG_{1,\rho_3,0,\rho_3}^2-\frac{5}{16}\cG_{1,\rho_3,1,1}^2\\
 \nonumber &+\frac{1}{16}\cG_{1,\rho_3,1,\rho_3}^2-\frac{1}{16}\cG_{\rho_2,0,1,0}^1+\frac{1}{16}\cG_{\rho_2,0,1,1}^1-\frac{1}{16}\cG_{\rho_2,0,1,\rho_2}^1-\frac{1}{16}\cG_{\rho_2,0,\rho_2,1}^1\\
 \nonumber &+\frac{1}{8}\cG_{\rho_2,1,0,0}^1-\frac{1}{16}\cG_{\rho_2,1,0,1}^1-\frac{1}{16}\cG_{\rho_2,1,0,\rho_2}^1-\frac{3}{16}\cG_{\rho_2,1,1,0}^1+\frac{1}{8}\cG_{\rho_2,1,1,1}^1\\
 \nonumber &+\frac{1}{8}\cG_{\rho_2,1,1,\rho_2}^1+\frac{1}{16}\cG_{\rho_2,1,\rho_2,1}^1+\frac{1}{8}\cG_{\rho_2,\rho_2,1,0}^1-\frac{1}{8}\cG_{\rho_2,\rho_2,1,1}^1-\frac{1}{16}\cG_{\rho_3,0,0,1}^2\\
 \nonumber &+\frac{1}{8}\cG_{\rho_3,0,1,1}^2-\frac{1}{16}\cG_{\rho_3,0,1,\rho_3}^2+\frac{1}{16}\cG_{\rho_3,0,\rho_3,1}^2+\frac{1}{16}\cG_{\rho_3,1,0,0}^2+\frac{1}{16}\cG_{\rho_3,1,0,1}^2\\
 \nonumber &-\frac{1}{16}\cG_{\rho_3,1,0,\rho_3}^2-\frac{3}{16}\cG_{\rho_3,1,1,1}^2+\frac{1}{8}\cG_{\rho_3,1,1,\rho_3}^2-\frac{1}{8}\cG_{\rho_3,1,\rho_3,0}^2+\frac{1}{16}\cG_{\rho_3,1,\rho_3,1}^2\\
 \nonumber &+\frac{1}{8}\cG_{\rho_3,\rho_3,1,0}^2-\frac{1}{8}\cG_{\rho_3,\rho_3,1,1}^2+\frac{1}{2}\cG_{0,1}^2\zeta_2+\frac{1}{2}\cG_{1,0}^2\zeta_2-\cG_{1,1}^2\zeta_2-\frac{1}{2}\zeta_3\cG_1^2\\
 \nonumber &+\frac{3}{16}\cG_{0,0}^2\cG_{0,1}^1+\frac{1}{8}\cG_{0,0}^3\cG_{0,1}^2+\frac{1}{16}\cG_{0,1}^2\cG_{0,1}^3-\frac{3}{16}\cG_{0,1}^3\cG_{0,\rho_3}^2-\frac{3}{16}\cG_{0,0}^2\cG_{1,0}^1\\
 \nonumber &+\frac{1}{8}\cG_{0,1}^2\cG_{1,0}^1+\frac{1}{8}\cG_{0,0}^3\cG_{1,0}^2-\frac{1}{8}\cG_{0,1}^1\cG_{1,0}^2-\frac{1}{16}\cG_{0,1}^3\cG_{1,0}^2+\frac{1}{16}\cG_{1,0}^2\cG_{0,\rho_2}^1\\
 \nonumber &+\frac{1}{8}\cG_{1,0}^1\cG_{1,0}^2-\frac{3}{16}\cG_{0,1}^2\cG_{1,0}^3+\frac{1}{16}\cG_{1,0}^3\cG_{0,\rho_3}^2-\frac{3}{16}\cG_{1,0}^2\cG_{1,0}^3+\frac{1}{8}\cG_{1,0}^2\cG_{1,1}^1\\
 \nonumber &-\frac{1}{4}\cG_{0,0}^3\cG_{1,1}^2-\frac{1}{16}\cG_{0,1}^1\cG_{1,1}^2+\frac{3}{16}\cG_{0,1}^3\cG_{1,1}^2-\frac{3}{16}\cG_{1,1}^2\cG_{0,\rho_2}^1-\frac{1}{16}\cG_{1,0}^1\cG_{1,1}^2\\
 \nonumber &+\frac{3}{16}\cG_{1,0}^3\cG_{1,1}^2-\frac{1}{8}\cG_{1,1}^1\cG_{1,1}^2+\frac{1}{4}\cG_{1,1}^3\cG_{0,\rho_3}^2+\frac{1}{8}\cG_{1,0}^2\cG_{1,1}^3-\frac{1}{8}\cG_{1,1}^2\cG_{1,1}^3\\
 \nonumber &+\frac{3}{16}\cG_{0,0}^2\cG_{1,\rho_2}^1-\frac{5}{16}\cG_{0,1}^2\cG_{1,\rho_2}^1-\frac{3}{16}\cG_{1,0}^2\cG_{1,\rho_2}^1+\frac{7}{16}\cG_{1,1}^2\cG_{1,\rho_2}^1-\frac{3}{16}\cG_{0,0}^3\cG_{1,\rho_3}^2\\
 \nonumber &+\frac{3}{16}\cG_{0,1}^3\cG_{1,\rho_3}^2+\frac{1}{8}\cG_{1,0}^3\cG_{1,\rho_3}^2-\frac{1}{4}\cG_{1,1}^3\cG_{1,\rho_3}^2+\frac{1}{8}\cG_{0,1}^2\cG_{\rho_2,0}^1+\frac{1}{16}\cG_{1,0}^2\cG_{\rho_2,0}^1
 \end{align}
 \begin{align}
 \nonumber \phantom{\tilde{g}_{++++}^{(0,1,0)} \left( \rho_1, \rho_2, \rho_3 \right) } &-\frac{3}{16}\cG_{1,1}^2\cG_{\rho_2,0}^1-\frac{1}{16}\cG_{0,0}^2\cG_{\rho_2,1}^1-\frac{3}{16}\cG_{0,1}^2\cG_{\rho_2,1}^1-\frac{1}{16}\cG_{1,0}^2\cG_{\rho_2,1}^1+\frac{5}{16}\cG_{1,1}^2\cG_{\rho_2,1}^1\\
 \nonumber &+\frac{1}{8}\cG_{0,1}^2\cG_{\rho_2,\rho_2}^1-\frac{1}{8}\cG_{1,1}^2\cG_{\rho_2,\rho_2}^1-\frac{1}{16}\cG_{0,1}^3\cG_{\rho_3,0}^2+\frac{1}{16}\cG_{1,0}^3\cG_{\rho_3,0}^2+\frac{1}{8}\cG_{1,1}^3\cG_{\rho_3,0}^2\\
 \nonumber &+\frac{1}{16}\cG_{0,0}^3\cG_{\rho_3,1}^2-\frac{1}{8}\cG_{0,1}^3\cG_{\rho_3,1}^2-\frac{1}{16}\cG_{1,0}^3\cG_{\rho_3,1}^2+\frac{1}{8}\cG_{0,1}^3\cG_{\rho_3,\rho_3}^2-\frac{1}{8}\cG_{1,1}^3\cG_{\rho_3,\rho_3}^2\\
 \nonumber &+\frac{3}{16}\cG_1^1\cG_{0,0,0}^2-\frac{3}{16}\cG_1^2\cG_{0,0,0}^3+\frac{1}{8}\cG_0^2\cG_{0,0,1}^1-\frac{1}{8}\cG_1^2\cG_{0,0,1}^1+\frac{1}{16}\cG_0^3\cG_{0,0,1}^2\\
 \nonumber &-\frac{1}{8}\cG_1^1\cG_{0,0,1}^2-\frac{1}{16}\cG_1^3\cG_{0,0,1}^2-\frac{1}{16}\cG_{0,0,1}^2\cG_{\rho_2}^1+\frac{1}{8}\cG_{0,0,1}^3\cG_{\rho_3}^2-\frac{1}{8}\cG_1^2\cG_{0,0,\rho_2}^1\\
 \nonumber &+\frac{1}{8}\cG_1^3\cG_{0,0,\rho_3}^2-\frac{1}{8}\cG_0^2\cG_{0,1,0}^1+\frac{1}{8}\cG_1^2\cG_{0,1,0}^1-\frac{1}{16}\cG_1^1\cG_{0,1,0}^2-\frac{1}{16}\cG_{0,1,0}^2\cG_{\rho_2}^1\\
 \nonumber &+\frac{1}{4}\cG_1^2\cG_{0,1,0}^3-\frac{1}{16}\cG_{0,1,0}^3\cG_{\rho_3}^2-\frac{3}{16}\cG_0^2\cG_{0,1,1}^1+\frac{3}{16}\cG_1^2\cG_{0,1,1}^1+\frac{1}{16}\cG_1^1\cG_{0,1,1}^2\\
 \nonumber &+\frac{1}{16}\cG_{0,1,1}^2\cG_{\rho_2}^1+\frac{1}{16}\cG_1^2\cG_{0,1,1}^3-\frac{3}{16}\cG_{0,1,1}^3\cG_{\rho_3}^2+\frac{3}{16}\cG_0^2\cG_{0,1,\rho_2}^1+\frac{1}{8}\cG_1^2\cG_{0,1,\rho_2}^1\\
 \nonumber &+\frac{1}{8}\cG_0^3\cG_{0,1,\rho_3}^2-\frac{1}{4}\cG_1^3\cG_{0,1,\rho_3}^2+\frac{1}{8}\cG_1^2\cG_{0,\rho_2,0}^1+\frac{1}{16}\cG_1^2\cG_{0,\rho_2,1}^1+\frac{1}{8}\cG_1^3\cG_{0,\rho_3,0}^2\\
 \nonumber &-\frac{1}{8}\cG_0^3\cG_{0,\rho_3,1}^2-\frac{1}{4}\cG_1^3\cG_{0,\rho_3,1}^2+\frac{1}{8}\cG_0^2\cG_{1,0,0}^1-\frac{1}{8}\cG_1^2\cG_{1,0,0}^1-\frac{1}{16}\cG_0^3\cG_{1,0,0}^2\\
 \nonumber &+\frac{1}{16}\cG_1^3\cG_{1,0,0}^2-\frac{1}{16}\cG_{1,0,0}^2\cG_{\rho_2}^1+\frac{1}{8}\cG_1^2\cG_{1,0,0}^3-\frac{3}{16}\cG_0^2\cG_{1,0,1}^1+\frac{3}{16}\cG_1^2\cG_{1,0,1}^1\\
 \nonumber &+\frac{1}{8}\cG_0^3\cG_{1,0,1}^2-\frac{1}{16}\cG_1^1\cG_{1,0,1}^2-\frac{1}{8}\cG_1^3\cG_{1,0,1}^2+\frac{1}{8}\cG_{1,0,1}^2\cG_{\rho_2}^1-\frac{1}{16}\cG_1^2\cG_{1,0,1}^3\\
 \nonumber &-\frac{1}{16}\cG_{1,0,1}^3\cG_{\rho_3}^2+\frac{1}{16}\cG_1^2\cG_{1,0,\rho_2}^1-\frac{1}{16}\cG_0^3\cG_{1,0,\rho_3}^2-\frac{1}{8}\cG_1^3\cG_{1,0,\rho_3}^2-\frac{1}{16}\cG_0^2\cG_{1,1,0}^1\\
 \nonumber &+\frac{1}{16}\cG_1^2\cG_{1,1,0}^1+\frac{1}{4}\cG_0^3\cG_{1,1,0}^2-\frac{3}{16}\cG_1^1\cG_{1,1,0}^2-\frac{1}{4}\cG_1^3\cG_{1,1,0}^2+\frac{3}{16}\cG_{1,1,0}^2\cG_{\rho_2}^1\\
 \nonumber &-\frac{1}{16}\cG_1^2\cG_{1,1,0}^3+\frac{1}{16}\cG_{1,1,0}^3\cG_{\rho_3}^2+\frac{1}{8}\cG_0^2\cG_{1,1,1}^1-\frac{1}{8}\cG_1^2\cG_{1,1,1}^1-\frac{3}{16}\cG_0^3\cG_{1,1,1}^2\\
 \nonumber &+\frac{3}{16}\cG_1^1\cG_{1,1,1}^2+\frac{3}{16}\cG_1^3\cG_{1,1,1}^2-\frac{3}{16}\cG_{1,1,1}^2\cG_{\rho_2}^1-\frac{1}{8}\cG_1^2\cG_{1,1,1}^3+\frac{1}{8}\cG_{1,1,1}^3\cG_{\rho_3}^2\\
 \nonumber &-\frac{1}{4}\cG_1^2\cG_{1,1,\rho_2}^1-\frac{1}{4}\cG_0^3\cG_{1,1,\rho_3}^2+\frac{7}{16}\cG_1^3\cG_{1,1,\rho_3}^2-\frac{3}{16}\cG_0^2\cG_{1,\rho_2,0}^1+\frac{1}{4}\cG_1^2\cG_{1,\rho_2,0}^1\\
 \nonumber &+\frac{3}{16}\cG_0^2\cG_{1,\rho_2,1}^1-\frac{7}{16}\cG_1^2\cG_{1,\rho_2,1}^1+\frac{1}{8}\cG_0^3\cG_{1,\rho_3,0}^2-\frac{1}{8}\cG_1^3\cG_{1,\rho_3,0}^2+\frac{1}{4}\cG_1^3\cG_{1,\rho_3,1}^2\\
 \nonumber &-\frac{1}{8}\cG_1^2\cG_{\rho_2,0,0}^1-\frac{1}{16}\cG_0^2\cG_{\rho_2,0,1}^1+\frac{1}{8}\cG_1^2\cG_{\rho_2,0,1}^1+\frac{1}{8}\cG_1^2\cG_{\rho_2,0,\rho_2}^1+\frac{3}{16}\cG_1^2\cG_{\rho_2,1,0}^1\\
 \nonumber &+\frac{1}{8}\cG_0^2\cG_{\rho_2,1,1}^1-\frac{1}{4}\cG_1^2\cG_{\rho_2,1,1}^1-\frac{1}{16}\cG_0^2\cG_{\rho_2,1,\rho_2}^1-\frac{1}{16}\cG_1^2\cG_{\rho_2,1,\rho_2}^1-\frac{1}{8}\cG_1^2\cG_{\rho_2,\rho_2,0}^1\\
 \nonumber &+\frac{1}{8}\cG_1^2\cG_{\rho_2,\rho_2,1}^1-\frac{1}{16}\cG_0^3\cG_{\rho_3,0,1}^2-\frac{1}{8}\cG_0^3\cG_{\rho_3,1,0}^2+\frac{1}{8}\cG_0^3\cG_{\rho_3,1,1}^2+\frac{1}{16}\cG_1^3\cG_{\rho_3,1,1}^2\\
 \nonumber &+\frac{1}{16}\cG_0^3\cG_{\rho_3,1,\rho_3}^2-\frac{1}{16}\cG_1^3\cG_{\rho_3,1,\rho_3}^2-\frac{1}{8}\cG_1^3\cG_{\rho_3,\rho_3,0}^2+\frac{1}{8}\cG_1^3\cG_{\rho_3,\rho_3,1}^2\,.
\end{align}

\newpage

\bibliographystyle{JHEP}
\bibliography{refs}

\end{document}